\documentclass[a4wide,11pt]{article}

\pdfoutput=1 

\usepackage[table]{xcolor}
\usepackage{colortbl}
\RequirePackage{ifpdf} 
\usepackage{amsmath,amsfonts} 
\usepackage{mathtools}
\usepackage{cases}
\usepackage[T1]{fontenc}
\usepackage{upquote,multirow,slashed}
\usepackage{authblk}
\usepackage{cite,appendix}
\usepackage{empheq}
\usepackage{lipsum}

\usepackage{jheppub}

\usepackage{pstricks}
\usepackage[final]{pdfpages} 
\usepackage{ifpdf} 
\usepackage{slashed}
\usepackage[normalem]{ulem}
\usepackage{color}
\usepackage{xcolor}
\definecolor{urlblue}{rgb}{0.2,0.4,0.7}
\definecolor{citegreen}{rgb}{0,0.6,0.2}
\definecolor{linkred}{rgb}{0.9,0.2,0.1}
\usepackage{hyperref}
\hypersetup{colorlinks=true, citecolor=citegreen, linkcolor=blue, urlcolor=blue}

\usepackage{graphics}
\usepackage{etoolbox} 
\usepackage{fixmath}
\usepackage{psfrag}
\usepackage{autobreak}
\usepackage{marginnote}
\usepackage{scalerel}
\usepackage{enumitem}
\usepackage{appendix}
\usepackage{caption} 
\usepackage{notoccite} 

\newcommand{\NOdisplay}[1]{ }

\def\MSbar{\overline{\mathrm{MS}}}

\def\g5{\gamma_5}
\def\oVFF{G_{5\mathrm{f\bar{f}}}}

\def\gfanchor{{\fontfamily{qcr}\selectfont
g5anchor\,}}

\def\epsF{\epsilon_{\scaleto{F}{3.5pt}}}
\def\tprime{\textquotesingle{}t}

\def\mf{{\scaleto{m}{5.5pt}}_{\scaleto{f}{5.5pt}}}
\def\yf{{\scaleto{y}{7.6pt}}_{\scaleto{f}{5.5pt}}}

\title{A Procedure \gfanchor to Anchor $\gamma_5$ in Feynman Diagrams for the Standard Model
}

\author{Long Chen}
\emailAdd{longchen@sdu.edu.cn}
\affiliation{School of Physics, Shandong University, Jinan, Shandong 250100, China}

\abstract{
We present a procedure \gfanchor to anchor $\gamma_5$ in the definition of a Dirac trace with $\gamma_5$ in Dimensional Regularization (DR) in Feynman diagrams for the Standard Model, based on a recent revision of the works by Kreimer, Gottlieb and Donohue. For each closed fermion chain with an odd number of primitive (i.e.~not-yet-clearly-defined) $\gamma_5$ in a given Feynman diagram, {\fontfamily{qcr}\selectfont g5anchor\,} returns a definite set of anchor points for $\gamma_5$, in terms of pairs of ordered fermion propagators; at each of these $\gamma_5$ anchor points a fixed expression in terms of the Levi-Civita tensor and elementary Dirac matrices will be inserted together with a sign determined by anticommutatively shifting all $\gamma_5$ from their original places (dictated by the Feynman rules) to this anchor point. The defining expressions for the cyclic $\gamma_5$-odd Dirac traces in DR associated with closed fermion chains in amplitudes, or more generally squared amplitudes, thus follow from this procedure, where the Levi-Civita tensors are not necessarily treated strictly in 4-dimensions. We propose utilizing this definition in practical perturbative calculations in the Standard Model at least to two-loop orders with the current implementation. Certain limitations and modifications of the KKS and/or the Kreimer scheme are addressed, as well as the possible caveats with \gfanchor.
}

\begin{document}

\allowdisplaybreaks[4]
\unitlength1cm
\maketitle
\flushbottom

\section{Introduction}
\label{sec:intro} 

The Standard Model (SM) for particle physics is a multiplicatively-renormalizable Quantum Field Theory based on the (chiral) gauge group $\mathrm{SU}_c(3) \otimes \mathrm{SU}_L(2) \otimes \mathrm{U_Y}(1)$ with spontaneous electroweak symmetry breaking by the non-zero vacuum condensation of the Higgs field.
Much of its predictive power lies in the celebrated multiplicative renormalizability, which among many other things ensures that the total number of parameters to be determined experimentally for this theory is fixed.\footnote{Effective field theories are practically renormalizable when truncated to fixed perturbative orders, and do retain predictive powers as we can have or find more experimental data than the number of parameters in the truncated effective Lagrangian (which, however, may increase along with the perturbative order).}
The renormalizability of a non-Abelian gauge theory is highly non-trivial to prove, and was addressed in many classical works~\cite{tHooft:1971akt,tHooft:1971qjg,tHooft:1972tcz,tHooft:1972qbu,Lee:1972fj,Lee:1972ocr,Lee:1972yfa,Weinberg:1971fb,Fujikawa:1972fe,Weinberg:1973ew,Weinberg:1973ua}, first without and later with spontaneous symmetry breaking.
What plays an essential role in the renormalizability, as well as unitarity, 
of a local gauge field theory are the Ward-Takahashi (WT) identities and the generalized ones in non-Abelian gauge theories known as Slavnov-Taylor (ST) identities, the relations among Green's correlation functions linked to the gauge symmetries~\cite{Becchi:1974xu,Becchi:1974md,Becchi:1975nq}. 
The formal derivation of various WTs in the SM can be systematically done by employing the notion of generating functionals for Green's correlation functions, which are well-established at the formal level. (See also, e.g.\,refs.~\cite{Aoki:1982ed,Bohm:1986rj,Denner:1994xt,Kraus:1997bi,Grassi:1999nb,Bohm:2001yx,Denner:2019vbn,Belusca-Maito:2023wah}.)

Renormalization in the electroweak (EW) sector of SM is much more involved than in the sector of Quantum Chromodynamics (QCD), both due to the mixing among gauge bosons of different gauge groups, and the chiral interactions with fermions.
Following the aforementioned classical works, the details of the renormalization counter-terms needed for the EW sector were worked out in refs.~\cite{Aoki:1982ed,Bohm:1986rj,Denner:1994xt,Bohm:2001yx,Denner:2019vbn} at one-loop level, and further at 2-loop level~\cite{Freitas:2002ja,Awramik:2002vu,Actis:2006ra,Actis:2006rb,Actis:2006rc}.\footnote{It is worth mentioning that the explicit analytical results for the beta functions for the three gauge couplings of the SM in the modified minimal-subtraction ($\MSbar$) scheme have been worked out to three loops, taking into account Yukawa and Higgs self-couplings~\cite{Mihaila:2012fm,Mihaila:2012pz,Bednyakov:2012rb}, and even to 4-loop order\cite{Martin:2015eia,Zoller:2015tha,Chetyrkin:2016ruf,Bednyakov:2015ooa,Davies:2019onf} (under reasonable approximations); limited to QCD, the state-of-the-art results for the $\alpha_s$ and mass anomalous dimension have reached 5-loop order \cite{Baikov:2016tgj,Herzog:2017ohr,Luthe:2017ttg,Baikov:2014qja,Luthe:2016xec,Baikov:2017ujl}.}
However, a subtle aspect in applying these renormalization findings to a general scattering process, which was acknowledged but not thoroughly addressed in those seminal papers (though notably underscored in ref.~\cite{Jegerlehner:2000dz}), concerns the very existence of a practical regularization prescription upholding all the defining fundamental Lorentz and chiral gauge symmetries in SM.~
Cancellation of axial or Adler-Bell-Jackiw (ABJ) anomalies~\cite{Adler:1969gk,Bell:1969ts} in all SM gauge currents, demonstrated explicitly e.g.~in refs.~\cite{Bouchiat:1972iq,Gross:1972pv,Geng:1989tcu}, is crucial in this regard. 
The issue shall thus be how to formulate a practical regularization prescription, by construction, satisfying the assumed conditions to achieve the neat multiplicative renormalization of SM, rather than questioning the formal derivation and the resulting form of WTs. 
~\\

Dimensional regularization (DR)~\cite{tHooft:1972tcz,Bollini:1972ui} is currently the method of choice for high-order perturbative calculations in SM, especially in the QCD sector. 
However, it was known ever since the beginning that there is a limitation on this regularization when there are chiral interactions in the theory that feature intrinsic 4-dimensional objects like $\g5$.
Quoting from the original ref.~\cite{tHooft:1972tcz} where an explicit constructive definition for $\g5$ in terms of the 4-dimensional Dirac matrices is employed (hence not fully anticommuting with all Dirac matrices):
``The usual ambiguity of the choice of integration variables is replaced in our formalism by the ambiguity of the \textit{location} of $\g5$ in the trace''. 
In fact, a Dirac algebra with a fully anticommuting $\gamma_5$ in a generic D$\,\neq 4$ dimensions contradicts with the non-vanishing value of the trace of the products of one $\gamma_5$ and four $\gamma$ matrices in 4 dimensions.
Notably, a naive use of an anticommuting $\gamma_5$ in DR, where the invariance of loop integrals under arbitrary loop-momentum shifts is ensured, leads to the absence of the ABJ anomaly~\cite{Adler:1969gk,Bell:1969ts} for anomalous axial currents.
Nevertheless, the anticommutativity of $\gamma_5$ is essential for the concept of chirality of spinors in 4 dimensions.
To overcome these technical issues, various $\gamma_5$ prescriptions in DR have been developed to handle specific cases so far encountered in practical applications. (See, e.g.~refs.~\cite{tHooft:1972tcz,Akyeampong:1973xi,Breitenlohner:1977hr,Breitenlohner:1975hg,Breitenlohner:1976te,Bardeen:1972vi,Chanowitz:1979zu,Gottlieb:1979ix,Siegel:1979wq,Fujii:1980yt,Bonneau:1980yb,Ovrut:1981ne,Espriu:1982bw,Thompson:1985uv,Abdelhafiz:1986jh,Buras:1989xd,Kreimer:1989ke,Korner:1991sx,Kreimer:1993bh,Larin:1991tj,Larin:1993tq,Trueman:1995ca,Chetyrkin:1997gb,Pernici:1999ga,Jegerlehner:2000dz,Wu:2002xa,Ma:2005md,Tsai:2009it,Mihaila:2012pz,Fazio:2014xea,Moch:2015usa,Porto:2017asd,Bruque:2018bmy,Gnendiger:2017rfh,Zerf:2019ynn,Lang:2021hnw,Cherchiglia:2021uce,Rosado:2023ist,OlgosoRuiz:2024dzq} for an incomplete list.) 
Fortunately, the issue is not very severe at the one-loop order, 
and a careful treatment based on a naive anticommuting $\gamma_5$ was shown to be sufficient for SM up to this loop order, which are nowadays fully automatized~\cite{Draggiotis:2009yb,Garzelli:2009is,Garzelli:2010qm,Shao:2011tg}.
However, a proper and systematic treatment of $\gamma_5$ in DR beyond one-loop order is highly non-trivial, and more work are needed to this end. 
Application to effective field theories with chiral local-composite operators in the effective Lagrangian can only become more tricky (see, e.g.~refs~\cite{Buras:1989xd,Dugan:1990df,Adam:1993uu,Herrlich:1994kh,Chetyrkin:1997gb,Wang:2017ijn,Ahmed:2020kme,DiNoi:2023ygk,Egner:2024azu}), expected simply based on the fact that the structure of UV divergences in an effective field theory is typically much more involved than a multiplicatively-renormalizable fundamental field theory.

Within the scope of DR for loop integrals, there are two popular strategies employed in the aforementioned references to tackle the $\g5$ issue in perturbative calculations.\footnote{In the highly-efficient automated approaches to compute the amplitudes based on multiple on-shell cut reconstruction~\cite{Bern:1994zx,Bern:1994cg,Bern:1997sc,Britto:2004nc,Bern:2007dw,Ossola:2006us,Ellis:2007br} with Dirac trace done in 4 dimensions, e.g.~\cite{Draggiotis:2009yb,Garzelli:2009is,Garzelli:2010qm,Shao:2011tg,Lang:2021hnw}, the involvements of the $\g5$-related symmetry-restoration terms are implicit and effectively hidden in the determination of the so-called rational terms, which requires one way or another the information on the \textit{properly renormalized} amplitudes.}
One may choose the location of a non-anticommuting $\g5$ in the fermion chain, (in the same sense as referred to in the above quotation from ref.~\cite{tHooft:1972tcz} and hereafter called the \textit{anchor point}\footnote{We have intentionally avoided using the term ``reading-point'' in this article, as it had led to significant confusion in our discussions with colleagues, which mainly arose from its literal connection with trace cyclicity and potential preconceived meanings that differ from the intended interpretation.} for $\g5$ in the rest of the article,) to be the vertices dictated by the SM Feynman rules where an explicit definition for $\g5$ in terms of the 4-dimensional Levi-Civita tensor will be inserted. 
This kind of treatment is usually referred to as the Breitenlohner-Maison~\tprime~Hooft-Veltman (BMHV) scheme~\cite{tHooft:1972tcz,Breitenlohner:1977hr,Breitenlohner:1975hg,Breitenlohner:1976te}, or simply HV scheme for $\g5$ in short, and there are also variants~\cite{Akyeampong:1973xi,Larin:1991tj,Larin:1993tq} with special emphasis on the manifestly Hermitian form of the axial-current matrix. 
The most celebrated property of these schemes based on a constructively-defined non-anticommuting matrix is that the mathematical expression for any diagram with $\g5$ is unambiguously defined, in particular, irrespective of whether this object is embedded as a subgraph of a bigger diagram.
However, apart from leading to Dirac traces that are computationally challenging in the cases of multiple $\g5$ on the same fermion chain, another conceptually less favorable outcome of the loss of anticommutativity is that the WTs are not necessarily respected at the level of bare amplitudes in chiral gauge theories: 
additional spurious anomalous terms appear in the bare expressions of dimensionally regularized $\g5$-dependent diagrams with ultraviolet (UV) divergences~\cite{Bardeen:1972vi,Chanowitz:1979zu,Gottlieb:1979ix,Trueman:1979en,Fujii:1980yt,Bonneau:1980yb,Espriu:1982bw,Larin:1991tj,Larin:1993tq,Bos:1992nd}, which have to be removed order-by-order in the form of $\g5$-related symmetry-restoration renormalization.
Explicit perturbative results for these additional renormalizations are derived in refs.~\cite{Larin:1991tj,Larin:1993tq,Larin:1997qq,Rittinger:2012the,Ahmed:2021spj,Chen:2021gxv,Chen:2022lun} in QCD, for flavor non-singlet and singlet axial-current operators up to 5-loop order.
Application of a non-anticommuting $\gamma_5$ to a chiral gauge theory, in particular the SM, requires more counter-terms with new structures~\cite{Martin:1999cc,Grassi:1999tp,Sanchez-Ruiz:2002pcf,Shao:2011tg,Cornella:2022hkc,Belusca-Maito:2020ala,Belusca-Maito:2021lnk,Belusca-Maito:2023wah,Stockinger:2023ndm,OlgosoRuiz:2024dzq}, but could be carried out, in principle, to any perturbative order (See, e.g.~\cite{Breitenlohner:1977hr,Breitenlohner:1975hg,Breitenlohner:1976te,Martin:1999cc,Cornella:2022hkc,Belusca-Maito:2023wah}). 

In an alternative strategy, a judicious choice of the anchor point for $\g5$ is made, where $\g5$ is merely formally anticommuting as far as the shift from the original vertices in diagrams is concerned, in hope that WTs are respected automatically and then there is no more need to amend it manually~\cite{Bardeen:1972vi,Chanowitz:1979zu,Gottlieb:1979ix,Abdelhafiz:1986jh,Buras:1989xd,Kreimer:1989ke}.
A further promising development along this line was introduced in ref.~\cite{Korner:1991sx}, sometimes referred to as KKS scheme in literature, improving the treatment in the cases of fermion loops with an odd number of $\g5$, in the language of choosing the reading-points for the so-called ``non-cyclic trace''~\cite{Korner:1991sx,Kreimer:1993bh}. 
We followed up this route, in particular ref.~\cite{Kreimer:1993bh}, referred to as Kreimer scheme below, with insights from ref.~\cite{Gottlieb:1979ix}, but reformulated it in terms of familiar notions of standard (cyclic) Dirac traces and anchor points (specified in terms of a pair of incoming/outgoing fermion lines) that can be easily implemented with the publicly available computer algebra systems in ref.~\cite{Chen:2023lus}; 
we pointed out in addition that this prescription does not work as originally expected in ref.~\cite{Kreimer:1993bh} for amplitudes with \textit{external} axial anomaly: the ABJ anomaly equation does not hold automatically in the bare form, and the Adler-Bardeen theorem~\cite{Adler:1969er} is not observed, in this treatment.\footnote{As a kind reminder, the appearance of the (gauge-non-invariant) Chern-Simons current term in the renormalization of the singlet axial-current in QCD~\cite{Chen:2023lus} shall not be taken as a counter-example to the general theorem on renormalization, but rather viewed as manifestation of a technical regularization issue: if the dimensional regularization scheme in use violates certain gauge symmetry or 4-dimensional $\g5$-involved relations (e.g.~the ABJ equation), counter-terms may be added to restore those relations which are non-invariant in general (see, e.g.~\cite{Buras:1989xd,Cornella:2022hkc,Belusca-Maito:2023wah}).}
There was also an outline of how the procedure may be performed in an algorithmic manner beyond the QCD corrections in ref.~\cite{Chen:2023lus}, but no systematic formulation of the procedure was given explicitly nor was there a ready-to-use implementation requested later by readers with interest.
However, the treatment in the case of divergent phase-space integration was not elaborated. 
In this work, we fill this gap by providing the first proof-of-concept implementation of the procedure sketched in ref.~\cite{Chen:2023lus}, together with a more streamlined formulation.

There are several other aspects underlying the revision compared to the original formulation~\cite{Korner:1991sx,Kreimer:1993bh}. 
In particular, although this kind of treatment is expected to be applicable to the QCD corrections to the matrix elements of non-anomalous non-singlet axial-current operators, 
there is currently no consensus in the literature on its applicability  to electroweak corrections to all orders.
We will discuss some of the conditions implied in ref.~\cite{Kreimer:1993bh} that are, unfortunately, not quite satisfied in the full SM at sufficiently high orders.
In addition, potential differences may be entailed by the modifications concerning the exact choices of $\g5$ anchor points, the treatment of Levi-Civita tensors,\footnote{This may start to become relevant only for sufficiently involved graph topology. Unfortunately it is not absolutely clear to the author the exact meaning of the ``$A$ vertices'' and the role of pseudo-scalar vertices in ref.~\cite{Kreimer:1993bh} in the presence of multi-loop electroweak corrections in general. (There are also interpretations in the literature concerning this aspect different from ours.) On the other hand, for each given Feynman diagram in electroweak SM, our choices of anchor points for $\g5$ are those determined by the application of \gfanchor.} and a clear-cut recipe in the case of squared amplitudes to be integrated over phase-space with intermediate infrared divergences. 
~\\

The remainder of this article is organized as follows.
The prescription to define $\g5$-odd Dirac traces in DR using the standard Dirac algebra (albeit, maintaining $\g5$'s anticommutativity merely to a certain extent) is exposed in detail in the next section~\ref{sec:pres}.
In particular, a streamlined formulation of the technical procedure is presented in subsection~\ref{sec:pres_g5vertex}, applicable to the Case-$A$ diagrams defined later in subsection~\ref{sec:caseA} (where the  treatment of the Levi-Civita tensors in our $\g5$ prescription is also described).
In subsection~\ref{sec:epsion-MBA} we explain the reason why difficulties may be expected when applying the above procedure to the SM at high orders in general. 
In subsection~\ref{sec:caseB}, a particular example diagram from the Case-$B$ in SM is given where we suspect that the procedure may cease to work; 
subsequently we sketch a possible workaround to circumvent the difficulty in defining $\g5$-odd Dirac traces in Case-$B$ diagrams.
In subsection~\ref{sec:g5trace_squaredamps}, we discuss how to determine the $\g5$ anchor points for closed fermion chains in the so-called IR-correlated squared amplitudes, to avoid the introduction of spurious pieces in the final results for physical observables.
The last subsection~\ref{sec:tech} is devoted to a brief overview on the program \gfanchor where the procedure described in subsection~\ref{sec:pres_g5vertex} is implemented.
We conclude in section~\ref{sec:conc}.

\section{Prescription} 
\label{sec:pres} 

For the sake of readers' convenience, we start with recapitulating the basics underlying the $\g5$ prescription employed in this work.
It is essentially a reconstruction of the key messages revealed in ref.~\cite{Kreimer:1993bh} reformulated differently but completely in our understanding of the matter in view of the work~\cite{Gottlieb:1979ix}, apart from a few novel revision/extension alluded in the introduction.
Due to the subtlety surrounding the very existence of a Lorentz-covariant DR-based regularization scheme that manifestly preserves both the gauge and chiral symmetry at the bare level (i.e.~before resorting to the regulator's vanishing limit), we take a much more pragmatic and modest manner to introduce this matter: 
after stating what the expected conditions are, we describe how a $\g5$ prescription may be designed to have them fulfilled; 
whether this objective is achieved for SM at the perturbative order in question remains to be confirmed in principle.

\subsection{Treatment of fermion-boson vertices in SM}
\label{sec:pres_g5vertex}

From the SM Lagrangian in the manifestly-renormalizable $R_{\xi}$-gauge~\cite{tHooft:1972qbu,Fujikawa:1972fe}, the vertices through which $\g5$ appear in Feynman diagrams are the axial components of the fermion's coupling to EW gauge bosons, denoted schematically as $\bar{\psi} \gamma_{\mu}\g5 \psi A^{\mu}_i$, and the pseudo-scalar Yukawa couplings to the associated would-be Goldstone bosons denoted schematically as $\bar{\psi}i\g5 \psi \phi_i$.
Both $\bar{\psi} \gamma_{\mu}\g5 \psi A^{\mu}_i$ and $\bar{\psi}i\g5 \psi \phi_i$ have mass-dimension 4, the highest allowed in an all-order renormalizable theory. 
The corresponding amputated 1PI 3-point Green functions have a superficial UV degree zero (and do possess overall UV divergences), and any more external fields included on top of this configuration will lead to a negative superficial UV degree, hence absence of overall UV divergences. 

\begin{figure}[htbp] 
\begin{center}
\includegraphics[scale=0.30]{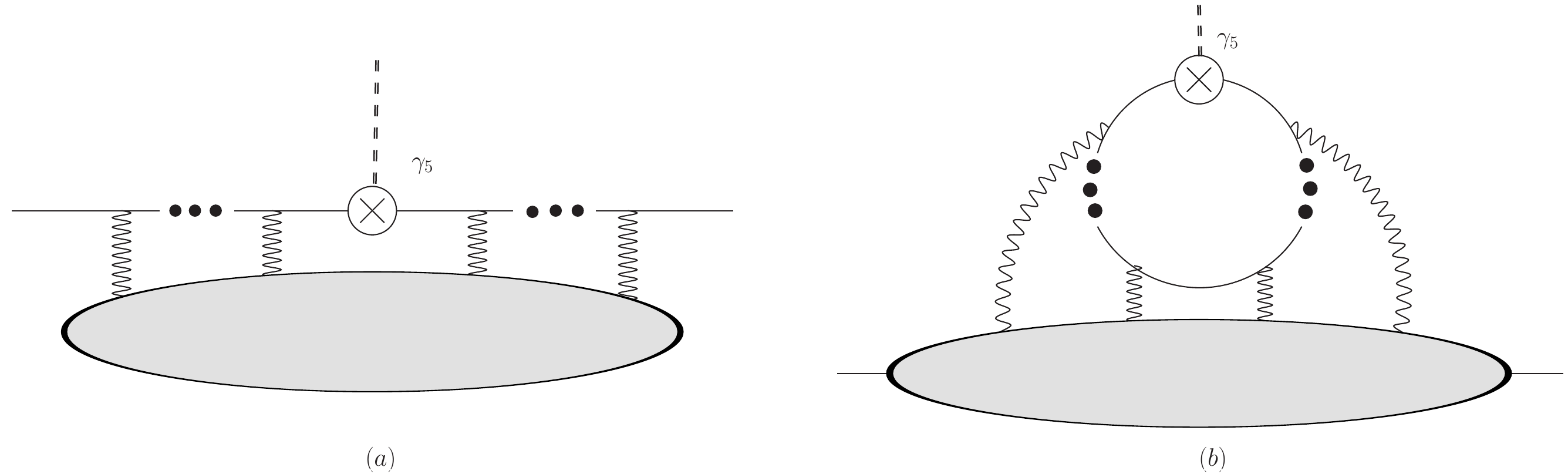}
\caption{An illustration of the OF-type (left, $a$) and CF-type (right, $b$) loop diagrams contributing to the 1PI vertex functions where the encircled cross denotes an insertion of either an external axial-current or a pseudo-scalar operator with a $\g5$ matrix which may be related to certain external EW boson denoted by the double-dashed line. The wavy lines denote gauge bosons of an Abelian or non-Abelian gauge theory, while the solid lines correspond to fermions. The grey blob and the dots on the solid fermion lines represent arbitrary additional virtual interactions not drawn explicitly.}
\label{fig:AxialVertexLoopGraphs}
\end{center}
\end{figure}

As in refs.~\cite{Adler:1969er,Adler:1969gk}, the loop corrections to the 1PI 3-point Green functions with one insertion of an external axial-current or pseudo-scalar operator between a pair of fermion fields can be organized into two categories, solely based on whether the axial-current or pseudo-Yukawa coupling in question is directly attached to the open fermion line running through the corresponding loop diagrams: 
if this is the case, such as shown in figure~\ref{fig:AxialVertexLoopGraphs}-(a), it will be called an OF-type correction; 
otherwise, namely if the $\g5$-vertex is attached to an internal closed fermion loop, such as shown in figure~\ref{fig:AxialVertexLoopGraphs}-(b), we call the correction CF-type.

For the OF-type loop corrections so identified, it is well-known~\cite{Bardeen:1972vi,Chanowitz:1979zu,Gottlieb:1979ix} how to manipulate $\g5$ in this type of dimensionally-regularized Feynman diagrams without spoiling the cherished Ward-Takahashi identities:
\begin{itemize}
\item 
(i) $\g5$ shall be pulled, fully anticommutatively according to
\begin{equation}
\label{eq:ACgamma5}
\{\gamma^{\mu},\, \gamma_5\} = 0 \,,
\end{equation}
outside the whole overall-divergent 1PI fermion-boson vertex, where $\g5^2=\hat{1}$ can be applied to eliminate $\g5$-pairs on the same fermion chain; 
\item 
(ii) after shifting $\g5$ anticommutatively to some chosen \textit{external} anchor point, the trace involving $\g5$ can be evaluated by inserting the following form in term of the fully-antisymmetric Levi-Civita tensor\footnote{We use the convention $\epsilon^{0123} = -\epsilon_{0123} = +1$. The treatment of $ \epsilon^{\mu\nu\rho\sigma}$, especially the contraction in pairs, will be discussed in detail in section~\ref{sec:LeviCiviaTensor}.},
\begin{equation}
\label{eq:gamma5}
\g5 \rightarrow -\frac{i}{4!}\epsilon^{\mu\nu\rho\sigma}\gamma_{\mu}\gamma_{\nu}\gamma_{\rho}\gamma_{\sigma} \equiv \hat{\gamma}_5
\end{equation}
which commutes with the loop integration in the conventional DR and also renormalization.
Note, however, the anticommutativity of the formal object $\g5$ is ``spontaneously'' lost upon substituting the expression~\eqref{eq:gamma5}, for which a specific notation $\hat{\gamma}_5$ is introduced for distinction.
(Because of this, and also the recipe to determine the anchor point below, might be more appropriate to describe the $\g5$ in this prescription as \textit{pseudo-anticommuting}.)
\end{itemize}

Now the recursive nature of the renormalization procedure, e.g.~as demonstrated in the Bogoliubov-Parasiuk-Hepp-Zimmermann (BPHZ) subtraction, naturally requires that the expressions for all overall-divergent 1PI Green functions shall be evaluated in the same way and to the same expression, irrespective of whether they are embedded as subgraphs in some Feynman diagrams. 
A proper prescription for anchor points of $\g5$ is introduced precisely to maintain the point-(i) and point-(ii) while complying with the above  \textit{consistency condition} in the cases where the fermion chain with $\g5$ may be closed into a loop such as the CF-type loop corrections in figure~\ref{fig:AxialVertexLoopGraphs}-(b).

It is worthy to note that in this regard a non-anticommuting $\g5$ scheme, e.g.~\cite{tHooft:1972tcz,Breitenlohner:1977hr,Breitenlohner:1975hg,Breitenlohner:1976te,Akyeampong:1973xi,Larin:1991tj,Larin:1993tq}, where every $\g5$ is anchored exactly where they were introduced by the Feynman rules in the Feynman diagrams, does have the above consistency condition automatically ensured; 
however, the defining WT/ST-identities of the SM are unfortunately not respected at the bare level and have to be restored by manually incorporating symmetry-restoration terms~\cite{Bardeen:1972vi,Chanowitz:1979zu,Gottlieb:1979ix,Trueman:1979en,Fujii:1980yt,Bonneau:1980yb,Espriu:1982bw,Larin:1991tj,Larin:1993tq,Bos:1992nd}.

In the BMHV scheme and its variants with a non-anticommuting $\g5$, to make the axial-current operator $\bar{\psi} \gamma_{\mu}\g5 \psi$ Hermitian as well as exhibiting the expected behavior under the charge-conjugation operation, the following ``anti-symmetric average'' shall be performed for the corresponding axial-current matrix~\cite{Akyeampong:1973xi,Fujii:1980yt,Collins:1984xc,Larin:1991tj,Larin:1993tq}:
\begin{equation}
\label{eq:NACg5Axialcurrent}
\gamma^{\mu}\gamma_5 \,\rightarrow \, 
\frac{1}{2} \big(\gamma^{\mu}\hat{\gamma}_5\, -\, \hat{\gamma}_5\gamma^{\mu}\big)
= 
\frac{-i}{3!} \epsilon^{\mu\nu\rho\sigma} \gamma_{\nu} \gamma_{\rho} \gamma_{\sigma}    
\end{equation}
where, with a slight abuse of notation, $\hat{\gamma}_5$ in \eqref{eq:gamma5} was used in the r.h.s.~of the above replacement. 
Note that the equal sign in \eqref{eq:NACg5Axialcurrent} is an exact identity following just from the Dirac algebra $\{\gamma^{\mu},\, \gamma^{\nu}\} = 2\, g^{\mu\nu}$ and full antisymmetry property of $\epsilon^{\mu\nu\rho\sigma}$ (without assuming dimensionality of the Lorentz vector indices).
One can check that in absence of anticommutativity of $\gamma_5$, the anti-symmetric average in \eqref{eq:NACg5Axialcurrent} is needed to ensure the validity of the extended Furry's theorem (i.e.~with axial currents included) for fermion loops at the bare level.
~\\

It is also well-known~\cite{Bardeen:1972vi,Chanowitz:1979zu,Gottlieb:1979ix} that in the case of an even number of $\g5$ on the same fermion loop, \eqref{eq:ACgamma5} can be applied in DR along with the relation $\g5^2 = \hat{1}$, resulting in a \textit{unique} trace expression free of Levi-Civita tensors.
The independence of the resulting trace for a closed fermion chain with an even number of $\g5$ on the choice of locations or anchor points to which the pairs of $\g5$ are anticommuted and annihilated using $\g5^2 = \hat{1}$ can be appreciated in the following way~\cite{Chen:2023lus}. 
For each pair of $\g5$ on the closed fermion chain (in absence of Yukawa couplings), by the virtue of $\big(\slashed{p} + m \big)\, \gamma^{\mu} \, \g5 = \g5 \, \big(\slashed{p} -m \big)\, \gamma^{\mu}$ with an anticommuting $\g5$, there will be, at most, two symbolically different expressions generated by different choices of $\g5$ anchor points; they are related to each other by flipping the signs of all fermion-propagator masses.
If one considers gauge interactions as well as Yukawa couplings correlated with fermion masses as in SM (at least in sign), then the terms odd under a homogeneous sign-flip of all fermion-propagator masses do not contribute due to the vanishing traces of an odd number of Dirac-$\gamma$ matrices.
Consequently, the two aforementioned expressions are algebraically equivalent for closed fermion chains with an even number of $\g5$ (regardless of being on-shell cut or not).
Therefore, in practice there is no need to specify any specific $\g5$ anchor point for fermion loops with even number of $\g5$. 
Consequently, a prescription for non-trivial anchor points may be needed only for $\g5$-odd fermion loops.
~\\

\begin{figure}[htbp]
\begin{center}
\includegraphics[scale=0.90]{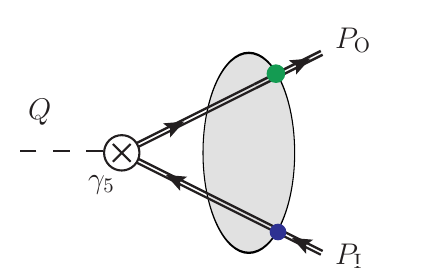}
\caption{An illustration of the maximal 1PI OF-type 3-point vertex graph $\oVFF$ where the $\g5$ matrix from an external axial or pseudo-scalar coupling vertex (denoted by the circle with a cross) lies on this open continuous fermion chain indicated by the double solid line. 
The grey blob denotes all 1PI OF-type loop corrections which may contain $\g5$ as well.
The arrows on the double lines represent the direction of the fermion-charge flow, which enters through the I-leg fermion propagator with incoming momentum $P_{\mathrm{I}}$ and leaves out via the O-leg fermion propagator with outgoing momentum $P_{\mathrm{O}}$ (subject to the momentum conservation $Q = P_{\mathrm{O}} - P_{\mathrm{I}}$). 
}
\label{fig:Max1PIopenVFF}
\end{center}
\end{figure}

The above information, especially point-(i,ii) and \eqref{eq:NACg5Axialcurrent}, together with the consistency condition for renormalization strongly suggests a blueprint for how to deal with $\g5$-odd fermion loops in a way with the anticommutativity~\eqref{eq:ACgamma5} preserved formally as much as possible. 
As far as UV renormalization is concerned, the most relevant loop corrections are OF-type 1PI graphs with exact two external fermion legs, which has a superficial UV-degree 0. 
(In SM any 1PI amplitude with additional external fermions or bosons has negative superficial UV-degree.)
Furthermore, the maximal one of this type, i.e.~the maximal 1PI OF-type  vertex correction (MIOFV), shall be naturally searched for as indicated in point-(i,ii), which is illustrated in figure~\ref{fig:Max1PIopenVFF} and denoted by $\oVFF$ for the sake of later reference. 
This is essentially figure~\ref{fig:AxialVertexLoopGraphs}-(a) but without requiring external fermions on-shell, and can appear as a subgraph in figure~\ref{fig:AxialVertexLoopGraphs}-(b).
We provided a simple procedure in ref.~\cite{Chen:2023lus} to identify the MIOFV sub-graph $\oVFF$ of a given fermion loop corresponding to a given external momentum insertion $Q$.
For the sake of reader's convenience, it is streamlined below, and implemented in \gfanchor presented in section~\ref{sec:tech}.
~\\

Let us consider a Feynman diagram $\mathrm{\mathbf{G}}$ containing the $\g5$-odd fermion loop $F_G$ of interest, whose closed fermion chain is denoted as $F_c$.
$F_c$ can be written out in the direction \textit{against} the fermion-charge flow as usual, but is otherwise allowed to start from \textit{any} vertex or propagator chosen possibly randomly by the diagram generator in use, e.g.~QGRAF~\cite{Nogueira:1991ex}.
The symbolic expression for $F_c$ derived directly from the Feynman rules should be viewed as a \textit{bookkeeping} form, with the precise definition of the corresponding $\g5$-odd Dirac trace provided by the following steps.
\begin{itemize}
\item 
First of all, based on the graphical information of the original input $\mathrm{\mathbf{G}}$, the set of external momenta $E_Q$ of the target $\g5$-odd fermion loop $F_G$ may be identified by searching for the cut through a \textit{minimal} number of boson propagators of $\mathrm{\mathbf{G}}$ to \textit{isolate} $F_G$ into a 1PI-diagram $G$ that contains no other $\g5$-odd fermion loops\footnote{As will be clear in later subsections, this condition amounts to assuming that the diagram in question is of Case-$A$ type defined in subsection~\ref{sec:caseA}.} and has each of its external momenta equal to the difference between the momenta of certain pair of fermion propagators of $F_G$.
The subgraph $G$ could be equal to $\mathrm{\mathbf{G}}$.
(The details on how this is achieved can be found in the implementation \gfanchor made publicly available with this work.) 

\item  
Then, for each different external momentum $Q \in E_Q$ without linear degeneracy in kinematics (i.e.~none of its proper subsets satisfying the momentum conservation), find the corresponding subgraph $\oVFF$ with an open-fermion segment from $F_G$ as illustrated in figure~\ref{fig:Max1PIopenVFF}. 
The fermion leg of $\oVFF$ with incoming fermion-charge flow is marked as I-leg and the corresponding fermion propagator reads as $S_F^{\mathrm{I}}(P_{\mathrm{I}}) = \frac{i}{\slashed{P}_{\mathrm{I}} - m + i\epsF}$. 
Similarly, the other fermion leg of $\oVFF$ with outgoing fermion flow is marked as O-leg and the corresponding fermion propagator $S_F^{\mathrm{O}}(P_{\mathrm{O}})$. 
$\oVFF$ can be determined by examining all possible \textit{two-fermion cuts} through the fermion chain $F_c$, under the condition $P_{\mathrm{O}} -  P_{\mathrm{I}}$ equal to the momentum insertion $Q$, and then selecting the one resulting in the largest 1PI subgraph (containing the boson line carrying $Q$).
Limited to the 3-point graph $\oVFF$, the 1PI condition simply implies the absence of tree-propagators \textit{inside} $\oVFF$ with momenta either $P_{\mathrm{I}}$ or $P_{\mathrm{O}}$, in other words, $\oVFF$ is free of self-energy corrections to its external legs. 

The above is not applicable if $E_Q$ contains only one independent momentum, namely $F_G$ has exactly two boson legs, because the two $\oVFF$ graphs associated with the two legs overlap completely with each other.
Fortunately, any uncut 2-point $\g5$-odd fermion loop vanishes simply due to the full antisymmetry of $\epsilon^{\mu\nu\rho\sigma}$ (See subsection~\ref{sec:g5trace_squaredamps} for more discussions).

\item 
In view of point-(i), the two ending nodes of 
$\oVFF$, highlighted with color in figure~\ref{fig:Max1PIopenVFF}, constitute the feasible candidates of $\g5$ anchor points in this overall-divergent $\oVFF$, 
which are however not identical in general: 
the difference may not be algebraically zero in case of Lorentz indices of $\gamma^{\mu}$ being D$\,\neq 4$ dimensional, but does vanish once 4-dimensional constraints are taken into account (where the full anticommutativity of $\g5$ holds exactly in the  algebra in use).
\footnote{For OF-type amplitudes with additional real radiations, the possible difference generated between taking the external fermion legs of the whole diagram and those of the $\oVFF$ as the $\g5$ anchor points will become relevant only if these amplitudes contain infrared divergences in addition to the UV divergences considered so far. A consistent treatment of $\g5$ anchor points among a set of the so-called IR-correlated contributions leading to an IR-finite quantity shall be ensured at the level of squared amplitudes (where fermion chains are effectively closed). This will be discussed in detail in subsection~\ref{sec:g5trace_squaredamps}.}
In light of \eqref{eq:NACg5Axialcurrent} in BMHV scheme, an average between the expressions corresponding to these two $\g5$ anchor points, located at the two ends of the open fermion line of $\oVFF$, can be employed to ensure the extended Furry's theorem in this treatment of $\g5$-odd fermion loops in the same way.
This is why such kind of an operation was proposed in ref.~\cite{Kreimer:1993bh}, albeit in term of the notion of "non-cyclic trace"~\cite{Kreimer:1989ke,Korner:1991sx}.
To be more explicit, as a close analogue and generalization of \eqref{eq:NACg5Axialcurrent}, the expression $\bar{F}_c$ for the fermion chain $F_c$ is \textit{defined} as the following average:
\begin{equation}
\label{eq:FCdef}
\bar{F}_c \equiv \frac{1}{2}\, \Big(F^{\g5 \rightarrow \mathrm{Ih}}_c + F^{\g5 \rightarrow \mathrm{Ot}}_c \Big)
\end{equation}
where $F^{\g5 \rightarrow \mathrm{Ih}}_c$ denotes a definite string of Dirac $\gamma$-matrices obtained from the original bookkeeping form $F_c$ by anticommutatively shifting $\gamma_5$ from the original vertex to the \textit{head} of the I-leg propagator $S_F^{\mathrm{I}}(P_{\mathrm{I}})$, shown clearly in figure~\ref{fig:Max1PIopenVFF}, which is subsequently replaced by the constructive expression~\eqref{eq:gamma5}.
(Similarly, $F^{\g5 \rightarrow \mathrm{Ot}}_c$ denotes a definite string of $\gamma$-matrices obtained by inserting \eqref{eq:gamma5}, albeit at the \textit{tail} of $S_F^{\mathrm{O}}(P_{\mathrm{O}})$ in $F_c$ and taking into account the relative sign generated by anticommutatively shifting $\g5$ around.) 

\end{itemize}

As noted in point-(ii), upon the replacement~\eqref{eq:gamma5}, the anticommutativity of the original $\g5$ is lost, which results in a standard cyclic Dirac trace that can be read off starting from anywhere in the fermion chain.
After the Dirac algebra done according to the above prescription, 
the tensor loop integrals in $G$ are then defined and evaluated in Conventional DR~\cite{tHooft:1972tcz,Bollini:1972ui,Collins:1984xc} in the usual way. 
~\\

If the $\g5$-odd fermion loop in question is not inside any truly \textit{overall-divergent} 1PI diagram, then the treatment as described above and implemented in \gfanchor may be applicable. 
But can a $\g5$-odd fermion loop appear in an overall-divergent 1PI diagram in SM? 
Unfortunately, the answer is yes, although they only appear in a limited class of corrections beginning at sufficiently high loop orders to be discussed in the following. 

\subsection{Multi-boson amplitudes proportional to $\epsilon^{\mu\nu\rho\sigma}$ in SM: all finite but some not sufficiently soft}
\label{sec:epsion-MBA}

It is well-known that SM is free of the ABJ axial anomaly in all gauge currents, demonstrated explicitly e.g.~in refs.~\cite{Bouchiat:1972iq,Gross:1972pv,Geng:1989tcu}, which is crucial for the theoretical self-consistency of SM  
and in turn requires the SM leptons and quarks to appear in units of \textit{generations}. 
Following from the cancellation of axial gauge anomaly, the non-vanishing contribution from $\g5$-odd fermion loops to any multi-gauge-boson amplitude, stripping off all external polarization vectors, is proportional to the Levi-Civita tensor $\epsilon^{\mu\nu\rho\sigma}$ (referred below as ``$\epsilon^{\mu\nu\rho\sigma}$-dependent'' for short), and is free of \textit{overall} UV-divergence; 
hence they become UV finite after subtracting all possible sub-divergences. 
This shall be reflected in the absence of any $\epsilon^{\mu\nu\rho\sigma}$-dependent multi-gauge-boson counter-term in the SM Lagrangian in a regularization respecting the SM gauge symmetries.

\begin{figure}[htbp]
\begin{center}
\includegraphics[scale=0.50]{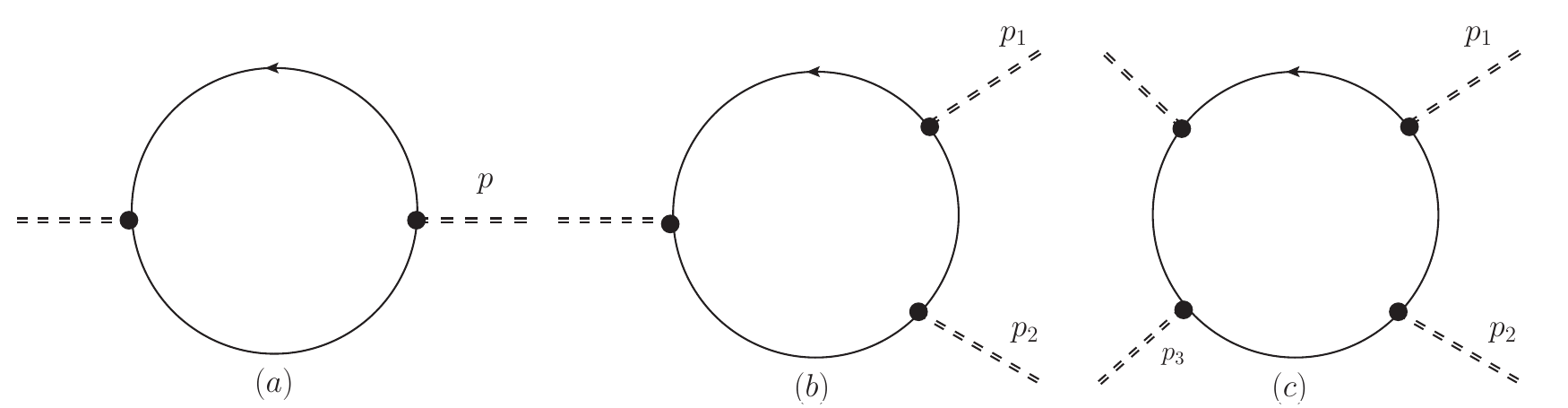}
\caption{An illustration of one-loop prototype diagrams for multi-boson amplitudes with non-negative superficial UV degree, that may have support for the structure $\epsilon^{\mu\nu\rho\sigma}$ (except for the first bubble graph (a) unless the fermion loop is cut open).
The solid circle with arrow denotes the $\g5$-odd fermion loop, and the external double-dashed lines are either gauge bosons or Higgs scalars with momenta labeled for the sake of reference (whose directions are irrelevant).
}
\label{fig:UVd1LMBgraphs}
\end{center}
\end{figure}

For $\epsilon^{\mu\nu\rho\sigma}$-dependent multi-boson amplitudes with external Higgs bosons (which could be the physical real Higgs boson or the unphysical would-be Goldstone bosons), since the Yukawa couplings to different SM fermions can be, a priori, arbitrarily different from each other, little can be inferred from the aforementioned cancellation of axial gauge anomaly. 
These amplitudes turn out to be free of overall-type UV divergence as well, simply owing to the properties of Dirac traces with $\gamma_5$ (see, e.g.~\cite{Mihaila:2012pz}).
To be more explicit, all possible relevant configurations with non-negative superficial UV degree are shown in figure~\ref{fig:UVd1LMBgraphs} at one-loop order. 
To distinguish from the notion of superficial UV degree in the following discussions, we call the actual leading scaling-power of a 1PI loop-amplitude in the overall-large loop momentum region as its \textit{proper UV degree}, that is after taking into account all possible cancellations due to certain gauge, Lorentz as well as charge-conjugation symmetries etc.

For the contribution from a $\g5$-odd (uncut) fermion loop to be non-zero, there must be at least 4 linearly independent external vectors 
involved in order to support a non-vanishing Levi-Civita tensor in the corresponding $\g5$-odd trace. 
Therefore a 2-point $\g5$-odd fermion loop shown in figure~\ref{fig:UVd1LMBgraphs}-(a) must vanish, unless being cut open with non-inclusive phase-space integration.
Consequently, a non-vanishing $\g5$-odd fermion loop starts from 3-point configuration, e.g.~figure~\ref{fig:UVd1LMBgraphs}-(b), where there must be at least two external or open Lorentz-vector indices. 
In figure~\ref{fig:UVd1LMBgraphs}-(b), there can be at most one external boson being a scalar, in which case the non-vanishing amplitude must be proportional to the Lorentz structure $\yf\, \mf \, \epsilon_{\mu_1\mu_2 \rho\sigma} \, p^{\rho}_1 \, p^{\sigma}_2$ with $\mf\,,\,\yf$ the mass and Yukawa coupling of the fermion in the loop.
The factor $\yf\, \mf$ appears due to the vanishing of Dirac traces with an odd number of Dirac $\gamma$-matrices. 
The mass-dimension of the Lorentz structure $\yf\, \mf \, \epsilon_{\mu_1\mu_2 \rho\sigma} \, p^{\rho}_1 \, p^{\sigma}_2$ is $+3$ and thus the proper UV degree of its coefficient must be $-2 < 0$ and hence free of overall UV divergence.
In case of one external momentum of figure~\ref{fig:UVd1LMBgraphs}-(b) vanishing, all three vertices must carry open Lorentz indices in order to have a support for $\epsilon^{\mu\nu\rho\sigma}$.
For figure~\ref{fig:UVd1LMBgraphs}-(c) with some external bosons being scalars, the possible non-vanishing Lorentz structures include $\yf\, \mf \, \epsilon_{\mu_1\mu_2 \mu_3 \rho} \, p^{\rho}_i$, $\yf^2 \, \epsilon_{\mu_1\mu_2 \rho \sigma} \, p^{\rho}_i p^{\sigma}_j$ and $\yf^3 \, m_f \, \epsilon_{\mu_1 \nu \rho \sigma} \, p^{\nu}_i p^{\rho}_j p^{\sigma}_k$, 
and their Lorentz invariant coefficients all have negative proper UV degree and hence free of overall UV divergences (which are UV finite).
Fermion loops with more than four fermion propagators become already superficially UV convergent.
~\\

Unfortunately, the above UV-finiteness alone is \textit{not sufficient} to ensure the absence of $\g5$-odd fermion loops in overall-divergent SM diagrams or amplitudes at high loop orders. 
To have the latter condition satisfied, these $\epsilon^{\mu\nu\rho\sigma}$-dependent multi-boson amplitudes shall be power-suppressed by the inverse of the large momenta $|p^{\mu}_i|$ in the limit $|p^{\mu}_i| \rightarrow \infty$, (i.e.~become soft) rather than approaching constants. 
One possible mechanism to ensure this is to demonstrate that these amplitudes feature some overall power factors in the fermion-propagator masses (hence vanish if all fermion propagators are massless) to provide the needed improvement in UV-power-counting when all momenta are large.
However, this is not the case for all $\epsilon^{\mu\nu\rho\sigma}$-dependent multi-boson amplitudes, with a notable exception being given by the 4-point figure~\ref{fig:UVd1LMBgraphs}-(c) with two external scalar bosons.\footnote{One of these two non-identical scalars shall couple to the fermion loop via pseudo-Yukawa coupling with $\g5$, otherwise, if $\g5$ comes from one of the two external vector currents there will be a cancellation between the two contributing diagrams with opposite fermion-charge flow on the fermion loop.
}
Once embedded into Feynman diagrams at sufficient high loop orders, subgrahs like this prototype eventually lead to some overall-divergent SM amplitudes containing $\g5$-odd fermion loops, contrary to what was expected in ref.~\cite{Kreimer:1993bh}. 
~\\

Let us now be more specific about this critical point. 
Firstly, upon cancellation or elimination of axial anomaly, the non-vanishing contributions from $\g5$-odd fermion triangle-loops necessarily depend on and are power-suppressed by the fermion-propagator masses~\cite{Adler:1969er,Adler:1969gk} (as they vanish in the limit of vanishing fermion mass); 
hence, these contributions do exhibit an improved power-counting behavior in the limit of large momentum flowing through the axial-current vertex, as analyzed and underscored particularly in ref~\cite{Gottlieb:1979ix} when discussing the treatment of $\g5$ in DR (albeit only for a limited class of diagrams).
\textit{Assuming} that the fermion-mass-independent parts of all $\epsilon^{\mu\nu\rho\sigma}$-dependent multi-gauge-boson amplitudes (not limited up to 4-point cases as shown in figure~\ref{fig:UVd1LMBgraphs}) are proportional to the \textit{anomaly-coefficients} of the gauge groups, then the non-vanishing contributions from all $\g5$-odd fermion-loops in SM shall feature at least an overall power-suppression factor quadratic in fermion masses. 
Under this \textit{assumed} condition, any $\g5$-odd fermion loop with only gauge-boson legs 
shall not appear inside any overall-divergent 1PI amplitude or Green function in the anomaly-free SM.\footnote{To this end, the massive gauge boson propagators are assumed to take their $R_{\xi}$-gauge~\cite{tHooft:1972qbu,Fujikawa:1972fe} form where they exhibit the usual inverse-square power counting behavior in the large momentum region.}

However, SM contains one more indispensable particle, the Higgs scalar, to generate masses for non-Abelian EW bosons and fermions without spoiling renormalizability and unitarity. 
Among its unique features relevant in our discussion here and below include that it interacts with massive SM fermions via Yukawa couplings that can be a priori arbitrarily different; 
furthermore, its self-energy function contains quadratic divergences and allows overall logarithmic UV divergences with coefficients quadratic in fermion-propagator masses.\footnote{Quadratic divergences manifest in DR as poles around $D=2$, which can be tracked if necessary.
}
It is crucial to note that for $\epsilon^{\mu\nu\rho\sigma}$-dependent multi-boson amplitudes with external scalars, the fermion-mass dependence from the mass-dimensionless Yukawa couplings does not necessarily lead to any improvement in the power-counting behavior in the large-momentum limit, only those picked up from the numerators of the fermion-loop propagators do. 
Despite the constraints from vanishing Dirac traces with odd $\gamma$-matrices and the extended Furry theorem with axial currents, one can indeed find concrete examples, in particular a box fermion loop as shown in figure~\ref{fig:UVd1LMBgraphs}-(c) with one scalar and one pseudo-scalar boson, where the Lorentz structure $\yf^2 \, \epsilon_{\mu_1\mu_2 \rho \sigma} \, p^{\rho}_i p^{\sigma}_j$ is not power-suppressed by the masses of fermion propagators. 
We have checked explicitly that the corresponding one-loop amplitudes are not vanishing with all fermion-propagators massless, but do vanish in the 4-dimensional limit if both scalars are identical.

Now one can compose non-vanishing multi-loop diagrams by gluing together a pair of subgraph~\ref{fig:UVd1LMBgraphs}-(c), which may lead to, for instance, a 2-point 4-loop diagram.
The non-trivial perturbative computations in refs.~\cite{Zoller:2015tha,Bednyakov:2015ooa} contained 4-loop diagrams exactly of this type, and the authors showed that they contribute $1/\epsilon$-pole UV divergences.
However, the existence of this type of contribution alone does not necessarily invalidate the Kreimer's prescription~\cite{Kreimer:1993bh}:
ref.~\cite{Bednyakov:2015ooa} pointed out explicitly the valid choices of "read-point(s)" that lead to the correct results~\cite{Poole:2019txl,Davies:2019onf}; 
and it is not completely inconceivable to refine the original $\g5$ prescription slightly such that only these valid choices are allowed. More comments on this are given in subsection~\ref{sec:pdexample}.
However, we will point out in the same subsection~\ref{sec:pdexample} that there are at least some other multi-loop diagrams involving the subgraph~\ref{fig:UVd1LMBgraphs}-(c) where there does not seem to exist any qualified choice to anchor $\g5$.
~\\

Before moving to the next subsection, we would like to demonstrate the following useful property for these overall-UV-finite objects:  
for the fermion-loop-induced contribution to an $\epsilon^{\mu\nu\rho\sigma}$-dependent multi-boson amplitude in SM, taking the divergence of an external axial-current vertex leads -- in case of employing formally anticommuting $\gamma_5$ -- to the appearance of power-suppression factors in fermion-propagator masses, regardless of whether there are external Higgs scalars.

\begin{figure}[htbp]
\begin{center}
\includegraphics[scale=0.30]{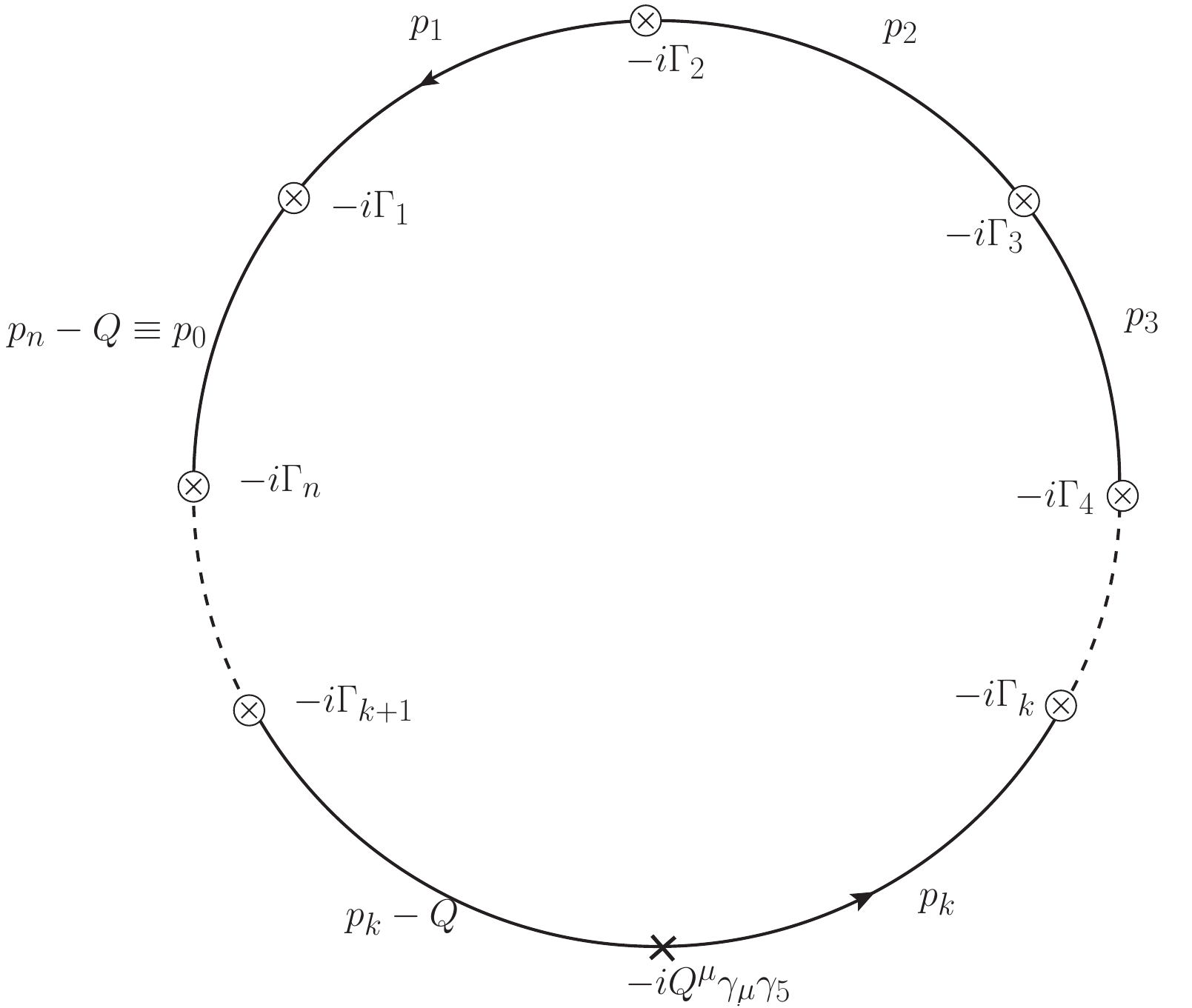}
\caption{A fermion loop with fermion propagators delimited by encircled crosses at which there may be some boson propagators attached, either external or internal, associated with certain vertex matrix $-i\Gamma_i$. 
The cross without circle between the $k$-th and $k+1$-th vertex denotes the insertion of the divergence of an external axial current with an incoming momentum $Q$.
The arrow on the solid line represents the direction of the fermion-charge flow on this fermion loop. The dashes between the two encircled crosses represent an indefinite number of fermion propagators not drawn explicitly.
The momentum assignments for fermion propagators are indicated explicitly on the diagram (along the direction of the fermion-charge flow). 
}
\label{fig:FermionLoop}
\end{center}
\end{figure}

Let us consider a prototype fermion loop with a kinematic configuration as illustrated in figure~\ref{fig:FermionLoop}.  
According to the momentum assignments for fermion propagators indicated explicitly there, the outgoing momentum through the $j$-th vertex reads 
\begin{eqnarray}\label{eq:FLkinematics}
q_j &\equiv& p_{j} - p_{j-1} \text{  for $j=1,2,\cdots, n$}   \nonumber\\ 
\sum_{j=1}^{n} q_i &=& \sum_{j=1}^{n} p_{j} - p_{j-1} = p_n - p_0 = Q\,,
\end{eqnarray}
with $p_0 \equiv p_n - Q$ parameterizing the loop momentum running in the fermion loop and $Q$ the outgoing momentum through the divergence of the external axial current in question (denoted by the cross without circle between the $k$-th and $k+1$-th vertex).
The $-i\Gamma_j$ retains only the Dirac matrices associated with the $j$-th vertex, with possible non-trivial color and charge factors stripped off. 
The Dirac matrix $\Gamma_j$ could be $\gamma^{\mu},\, \gamma^{\mu}\g5,\, \hat{1},\, \g5$ in our consideration below.
The Dirac trace corresponding to figure~\ref{fig:FermionLoop} reads
\begin{equation}\label{eq:FLtrace}
\mathrm{DT}_{[k]} \equiv \mathrm{Tr}\Big[ 
\big(\prod_{i=1}^{k} \Gamma_i \, S_F(p_i) \big) 
\big( \slashed{Q}\g5 \big) S_F(p_k - Q) 
\big(\prod_{j=k+1}^{n} \Gamma_j \, S_F(p_j - Q) \big)
\Big]
\end{equation}
with $S_F(p_i) = \frac{1}{\slashed{p}_i - m_i + i\epsF} = \frac{\slashed{p}_i + m_i }{p^2_i - m^2_i + i\epsF}$ denoting the fermion propagator with mass $m_i$.
For a given configuration of the original $n$ vertices, the total effect of exhausting all possible ways of insertion 
leads to a weighted sum $\sum_{k} \mathrm{cc}_{[k]} \, \mathrm{DT}_{[k]}$ where $\mathrm{cc}_{[k]}$ denotes the corresponding color/charge factor.
Keep in mind that the $\mathrm{DT}_{[k]}$ in \eqref{eq:FLtrace} always appears under the loop integration over $p_0$ done in DR, with $Q, q_1, q_2, \dots q_n$ fixed.

Below we first examine the simple scenario where the divergence of the axial current in question \textit{commutes} with the possible non-Abelian colors/charges carried by any of the original $n$ vertices. 
Consequently, for a given configuration of the original $n$ vertices, $\mathrm{cc}_{[k]}$ corresponding to different insertions are the same and can be pulled out from the sum $\sum_{k}  \mathrm{DT}_{[k]}$. 
This covers, for instance, the cases where the axial current to be inserted is diagonal in the space of fermion flavors, such as the one coupled to the Z boson, and all other vertices in figure~\ref{fig:FermionLoop} preserve flavors such as in the interactions with gluons, photon, Higgs and Z boson.

Now we employ the following rewriting  
\begin{eqnarray} \label{eq:pinchidentity}
\slashed{Q}\g5 &=& 
\big(\slashed{p}_k - m_i \big) \g5 + \g5 \big(\slashed{p}_k - \slashed{Q} - m_j \big)  + \big(m_i + m_j \big) \, \g5
\end{eqnarray}
which shall be applicable in our treatment of formally anticommuting $\g5$ in the anomaly-free SM.
Sandwiched between two neighboring fermion propagators in \eqref{eq:FLtrace}, we have 
\begin{eqnarray} \label{eq:pinchidentitySpecific}
\frac{1}{\slashed{p}_k - m_k} \Big(\slashed{Q}\g5 \Big)\frac{1}{\slashed{p}_k - \slashed{Q} - m_k}  
&=& 
\frac{1}{\slashed{p}_k - m_k} \Big(2\,m_k \g5  \Big)\frac{1}{\slashed{p}_k - \slashed{Q} - m_k} \nonumber\\
&+& \frac{1}{\slashed{p}_k - m_k} \g5 
+ \g5 \frac{1}{\slashed{p}_k - \slashed{Q} - m_k}  
\end{eqnarray}
The first $2\,m_k$-dependent term vanishes if all fermion propagators are massless, and clearly enjoys the property mentioned above.
What we would like to demonstrate explicitly below is that eventually the last two terms lead to contributions either vanishing or being power-suppressed by fermion-propagators' masses too, even if there are Yukawa vertices.

By \eqref{eq:pinchidentitySpecific}, $\mathrm{DT}_{[k]}$ can be rewritten as 
\begin{eqnarray}\label{eq:FLtraceRW}
\mathrm{DT}_{[k]} &=& \mathrm{DT}_{[k]}(\slashed{Q}\g5 \rightarrow (2\,m_k) \g5 ) \nonumber\\
&+&
\mathrm{Tr}\Big[ 
\big(\prod_{i=1}^{k-1} \Gamma_i \, S_F(p_i) \big) 
\big(  \Gamma_k \g5 \big) S_F(p_k - Q) 
\big(\prod_{j=k+1}^{n} \Gamma_j \, S_F(p_j - Q) \big)
\Big]\nonumber\\
&+& 
\mathrm{Tr}\Big[ 
\big(\prod_{i=1}^{k} \Gamma_i \, S_F(p_i) \big) 
\big(\g5 \Gamma_{k+1}  \big) S_F(p_{k+1} - Q) 
\big(\prod_{j=k+2}^{n} \Gamma_j \, S_F(p_j - Q) \big)
\Big]\nonumber\\
&\equiv& \mathrm{DT}^{ps}_{[k]} + \mathrm{DT}^{L}_{[k]} + \mathrm{DT}^{R}_{[k]}\,,
\end{eqnarray}
where in the last line three shorthand notations are introduced respectively for the three lines above.
Instead of being grouped in pairs 
as in $\sum_{k} \mathrm{DT}^{L}_{[k]} + \mathrm{DT}^{R}_{[k]}$ (with one common color/charge factor pulled out), the integrand terms of this dimensionally-regularized loop integration over $p_0$ can now be reorganized based on the vertex (labeled $K$ below) of which the insertion is made, respectively, to the l.h.s.~and r.h.s., as follows:%
\begin{eqnarray}\label{eq:FLtraceRW2}
\sum_{k} \mathrm{DT}^{L}_{[k]} + \mathrm{DT}^{R}_{[k]} \,\rightarrow\,  
\sum_{K} 
\mathrm{Tr}\Big[ 
\big(\prod_{i=1}^{K-1} \Gamma_i \, S_F(p_i) \big) 
\big( \Gamma_K \g5 + \g5 \Gamma_K \big) S_F(p_k - Q) 
\big(\prod_{j=K+1}^{n} \Gamma_j \, S_F(p_j - Q) \big)
\Big]\,,\nonumber\\
\end{eqnarray}
where the cyclicity of Dirac trace can be employed wherever needed to bring the l.h.s.~expression into this form.\footnote{The re-organization of the $n$ pairs of terms in $\sum_{k} \mathrm{DT}^{L}_{[k]} + \mathrm{DT}^{R}_{[k]}$ on the l.h.s.~of \eqref{eq:FLtraceRW2} is essentially to combine $\mathrm{DT}^{L}_{[k]}$ with $\mathrm{DT}^{R}_{[k-1]}$ into a single summand  (enumerated as the $K$-th piece in the sum) on the r.h.s.~of \eqref{eq:FLtraceRW2}.
}
So far no specific property of Dirac matrix $\Gamma_K$ was assumed, and the anticommutativity of $\g5$ was employed only in the perturbative identity~\eqref{eq:pinchidentity}.
We now consider the following two possible scenarios regarding $\Gamma_K$.
\begin{itemize}
 \item 
 If $\Gamma_K$ anti-commutes with $\g5$, i.e.~$\{\Gamma_K, ~\g5\} = 0$, such as when $\Gamma_K$ is a vector or axial-vector current matrix, 
 then $\sum_{k} \mathrm{DT}^{L}_{[k]} + \mathrm{DT}^{R}_{[k]}$ in \eqref{eq:FLtraceRW2} makes no contribution.
 \item 
 If $\Gamma_K$ commutes with $\g5$, i.e.~$[\Gamma_K, ~\g5] = 0$, such as when $\Gamma_K$ is a scalar or pseudo-scalar Yukawa-coupling matrix, 
 then $\big( \Gamma_K \g5 + \g5 \Gamma_K \big) = 2\,\g5 \Gamma_K$ and  \eqref{eq:FLtraceRW2} becomes 
\begin{eqnarray}\label{eq:FLtraceRW2yukawa}
\sum_{K} 
\mathrm{Tr}\Big[ 
\big(\prod_{i=1}^{K-1} \Gamma_i \, S_F(p_i) \big) 
\big( 2\,\g5 \Gamma_K \big) S_F(p_k - Q) 
\big(\prod_{j=K+1}^{n} \Gamma_j \, S_F(p_j - Q) \big)
\Big]\,.
\end{eqnarray}

There are now the following two possibilities:
 \begin{itemize}
 \item
 If the number of Yukawa vertices on the fermion loop is odd, the non-vanishing Dirac trace (with an even number of $\gamma$-matrices) in \eqref{eq:FLtrace}, without rewriting by \eqref{eq:pinchidentity}, will necessarily pick up at least one fermion-mass factor from the numerators of fermion propagators.

 \item
 If there are even number of Yukawa vertices on the fermion loop, there are no more overall fermion-mass power-factors in each individual Dirac traces in the sum~\eqref{eq:FLtraceRW2yukawa}.
 However, there then appears the following cancellation among the diagrams corresponding to different insertions of $\slashed{Q}\g5$: 
 say that there are two Yukawa vertices in the sum~\eqref{eq:FLtraceRW2yukawa}, the two Dirac traces corresponding to, respectively, having the insertion right next to them differ by a relative \textit{minus} sign in the limit of massless fermion propagators: the number of Dirac matrices for $\g5$ to be crossed over anticommutatively is odd if there are no other Yukawa vertices in-between;
 hence they cancel between each other, leading again to the vanishing of the sum \eqref{eq:FLtraceRW2yukawa} as a whole in this limit.

 \end{itemize}
\end{itemize}

When flavor-changing vertices appear on the fermion loop, namely including the full EW-gauge interactions, the commutation condition assumed for $\mathrm{cc}_{[k]}$ in the sum $\sum_{k} \mathrm{cc}_{[k]} \, \mathrm{DT}_{[k]}$ no longer holds.
There may appear additional terms with non-zero commutators involving EW-gauge group structures that are anti-symmetric when the charged axial-current operator is inserted to the left or right of a given vertex. 
It is not obvious, at least to the author, what would happen in general to these pieces 
after summing up all contributing diagrams with all possible permutations among the original $n$ vertices (including also those with reversed fermion-charge flow on the loop). 
Rather than proceeding further in the above diagrammatic way, we may take the following shortcut. 
\begin{itemize}
\item 
In the case of an odd number of Yukawa vertices on the fermion loop in figure~\ref{fig:FermionLoop}, irrespective of the properties of the other current-vertices, a non-vanishing Dirac trace must contain an even number of $\gamma$-matrices and thus necessarily picks up at least one mass factor from the fermion propagators' numerators, independent of whether the divergence of an axial current is taken.

\item 
In the less-transparent case with an even number of Yukawa vertices on the fermion loop, it might be convenient to directly appeal to a powerful theoretical asset available for the electroweak theory: the very relations~\cite{Lee:1977eg} or Ward identities~\cite{Chanowitz:1985hj} underlying the Goldstone boson equivalence theorem~\cite{Cornwall:1974km,Vayonakis:1976vz,Lee:1977eg,Chanowitz:1985hj}.
In particular, the discussion in the appendix of ref.~\cite{Lee:1977eg} is very illuminating: 
the gauge-fixing constraints for $\mathrm{SU}_{L}(2)$ EW-gauge fields $V^{\mu}_{a}(x)$ with mass $M_a$ (and $\phi_a(x)$ the corresponding would-be Goldstone boson) 
\begin{equation} \label{eq:thooftgaugefixing}
\partial_{\mu} V^{\mu}_{a}(x) -  M_a\, \phi_a(x) = 0
\end{equation} 
can be imposed as in the original \tprime~Hooft gauge~\cite{tHooft:1971qjg} in the perturbative calculations.\footnote{We note that perturbative calculations directly in the gauge-fixing \eqref{eq:thooftgaugefixing} are non-trivial in general, and we do not know any systematic demonstration or proof in literature that this can be done to arbitrary orders without any issue. 
This certainly poses as a limitation of the discussion presented here.} 
Furthermore, this condition shall hold, at each perturbative order, in the formal power expansion according to the number of fermion loops in the contributing diagrams.\footnote{If this had not been the case for SM, the concrete choices of the spinor-space dimension $\mathrm{Tr}[\hat{1}] = f(\epsilon)$ with $f(\epsilon \rightarrow 0) = 4$ in the definition of all Dirac traces in DR would affect, in a non-trivial way, the expressions of loop amplitudes; 
one would have trouble to justify the results derived under a particular choice, e.g.~the common setting $\mathrm{Tr}[\hat{1}] = 4$, to be ``physical'' when computing SM loop amplitudes in DR, as the expression would change had an alternative $f(\epsilon)$ been used.
}
Specialized into the considered class of contributions characterized by having an external massive EW gauge boson $V^{\mu}_{a}$ coupled to a closed fermion loop as in figure~\ref{fig:FermionLoop}, one thus expects that replacing the polarization vector of $V^{\mu}_{a}$ by its own momentum leads to expressions that would be the same as those obtained by replacing $\partial_{\mu} V^{\mu}_{a}$ by the corresponding would-be Goldstone boson $\phi_a$ (multiplied by $M_a$) that couples to the fermion loop by Yukawa interaction. 
The power-suppression by fermion-propagators' masses, which we are looking for in this case, then arises from the fact that with the appearance of $\phi_a(x)$ there are now an odd number of Yukawa vertices on the fermion loop in question. 
\end{itemize}

Therefore, for the fermion-loop induced contribution to an $\epsilon^{\mu\nu\rho\sigma}$-dependent multi-boson amplitude in SM, taking the divergence of an external axial-current vertex 
leads to the appearance of power-suppression factors in fermion-propagator masses, regardless of whether there are external Higgs scalars.
~\\

This property may be used to argue that the overall UV divergence associated with $\,\gamma_{\mu}\g5$ in the 3-point Green function between an axial-current operator and a pair of fermions does not receive contribution from the CF-type loop corrections which has the axial-current vertex on a $\g5$-odd fermion loop, as illustrated in figure~\ref{fig:AxialVertexLoopGraphs}-(b). 
(In contrast, we do not know any argument that can be used to exclude overall-divergent CF-type-alike loop corrections to the Yukawa interaction between a Higgs scalar and a pair of fermions that contain $\g5$-odd fermion loops.)
Note that to this end, we exploited also a special feature of this 3-point vertex function: 
there is just one Lorentz structure (i.e.$\,\gamma_{\mu}\g5$) with logarithmically overall-divergent Lorentz-invariant coefficient in its tensor decomposition and it is \textit{not} screened off when contracting with the external momentum flowing through the axial current.
This feature is, however, not typically present in the tensor decomposition of other more complicated Green functions, such as those with at least two external gauge bosons, hence similar statements could not be simply extended.

\subsection{Case-A: $\g5$-odd fermion loops not inside any overall-divergent 1PI amplitude}
\label{sec:caseA}

Even though the presence of overall-divergent 1PI diagrams with $\g5$-odd fermion loops is not excluded in SM as discussed above, it is straightforward to see that they appear only at sufficiently high loop orders. 

\subsubsection{A survey up to 3-loop orders in SM}

\begin{figure}[htbp]
\begin{center}
\includegraphics[scale=0.45]{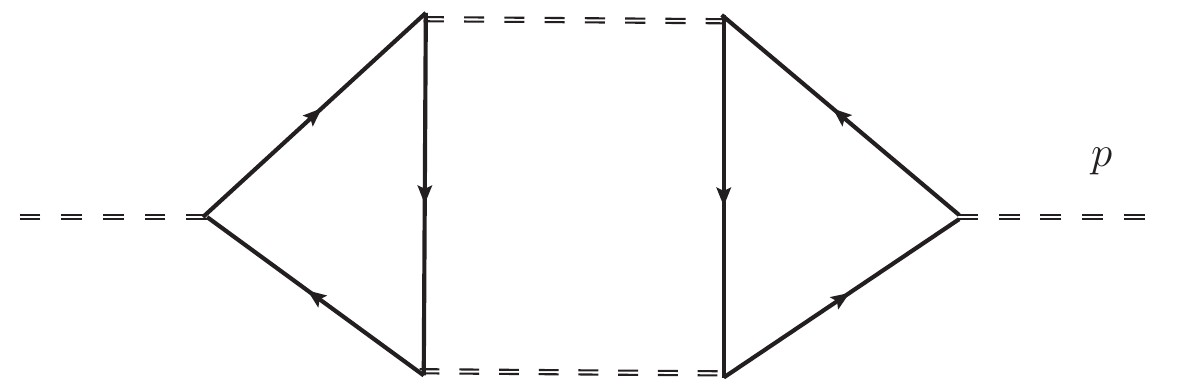}
\caption{
A prototype diagram for contributions to the self-energy functions of bosons with a pair $\g5$-odd fermion loops at 3-loop order. 
The solid triangles with arrow denote the $\g5$-odd fermion loops,
and the double-dashed lines are either gauge bosons or scalars. The external momentum flowing through the diagram is denoted as $p$.
}
\label{fig:DoubleTriangle}
\end{center}
\end{figure}

Let us first consider diagrams without open fermion chains, as will become more clear in subsection~\ref{sec:g5trace_squaredamps}, the $\g5$ anchor points need to be specified only for traces corresponding to closed fermion chains in amplitudes or squared amplitudes as far as computing physical observables are concerned.
According to the discussion at the beginning of section~\ref{sec:epsion-MBA},
to have non-vanishing support for $\epsilon^{\mu\nu\rho\sigma}$ from a $\g5$-odd fermion loop, there must be at least 4 linearly independent external vectors involved, and to this end the $\g5$-odd fermion loop shall have at least 3 vertices (two of which must carry open Lorentz vector indices). 
On the other hand, since all SM multi-boson amplitudes proportional to $\epsilon^{\mu\nu\rho\sigma}$ are devoid of overall UV-divergences, although some not sufficiently ``soft'' (i.e.~not sufficiently power-suppressed in the large-momentum limit), the potential occurrence of $\g5$-odd fermion loops in any overall-divergent 1PI multi-boson amplitude must occur in pairs, i.e.~in even numbers.
Therefore, these diagrams must start from 3-loop order, and the prototype for those with highest superficial UV-degree, hence the least number of external bosonic legs,   
is illustrated in figure~\ref{fig:DoubleTriangle}. 
In SM, all 3-loop corrections of this type~\ref{fig:DoubleTriangle} are vanishing in the limit of massless fermion propagators, either due to vanishing Dirac trace of odd numbers of $\gamma$ matrices and/or related to the cancellation of axial gauge anomaly (see, e.g.~\cite{Mihaila:2012pz}).

When the fermion propagators are massive, there can arise non-vanishing contributions from these 3-loop diagrams which must, however, contain power factors of fermion masses, as discussed in subsection~\ref{sec:epsion-MBA}.
In particular, for the cases with all double-dashed lines in  figure~\ref{fig:DoubleTriangle} being gauge bosons, this overall mass factor is at least of power 4 in fermion masses~\cite{Gottlieb:1979ix}, based on again the cancellation of gauge anomaly and properties of Dirac traces.
However, additional care is needed to demonstrate the UV-finitness of this class of 3-loop diagrams with two external scalar legs, where the superficial UV-degree is as high as $+2$, and moreover, overall UV divergences with fermion-mass-dependent prefactors are allowed in principle (i.e.~not excluded by gauge or chiral symmetry).
\begin{itemize}
    \item 
    If the two legs are both Higgs bosons, namely considering the self-energy function of Higgs boson, then the 3-loop diagrams with two $\g5$-odd fermion loops as in figure~\ref{fig:DoubleTriangle} make no contribution, due to the extended Furry theorem and properties of $\epsilon^{\mu\nu\rho\sigma}$.
    \item 
    In the case with two external pseudo-scalar bosons, combining the constraints from non-vanishing Dirac traces with $\epsilon^{\mu\nu\rho\sigma}$ from each of the two $\g5$-odd fermion loops, the non-vanishing terms require necessarily a non-vanishing external momentum $p$, and thus must be proportional to $m_{f_1}\, m_{f_2}\, p^2$ with $m_{f_1}\,, m_{f_2}$ masses of the fermion propagators; 
    this factor has a mass-dimension $+4$, and is enough to ensure a negative proper UV-degree and hence UV-finiteness of these 3-loop diagrams. 
\end{itemize}
Adding more external lines to the prototype graph in figure~\ref{fig:DoubleTriangle} will not increase the proper UV-degree\footnote{
For example, one may consider a 3-loop 4-point diagram with, e.g. two external
scalars and two external vectors, obtained by gluing together two 4-point fermion loops in figure~\ref{fig:UVd1LMBgraphs}-(c).
If all four external bosons' momenta are vanishing, then there is no more support for the Levi-Civita tensors in one of the 4-point fermion loops (each has only one independent momentum flowing through);
this means that at least one pair of the external bosons shall have non-vanishing external momenta for this diagram to be non-zero. 
Consequently, the form factors in its Lorentz tensor decomposition shall have negative mass dimensions.}, although introducing more internal lines (leading to more loops) may do as we will see in the next subsections.
~\\

\begin{figure}[htbp]
\begin{center}
\includegraphics[scale=0.48]{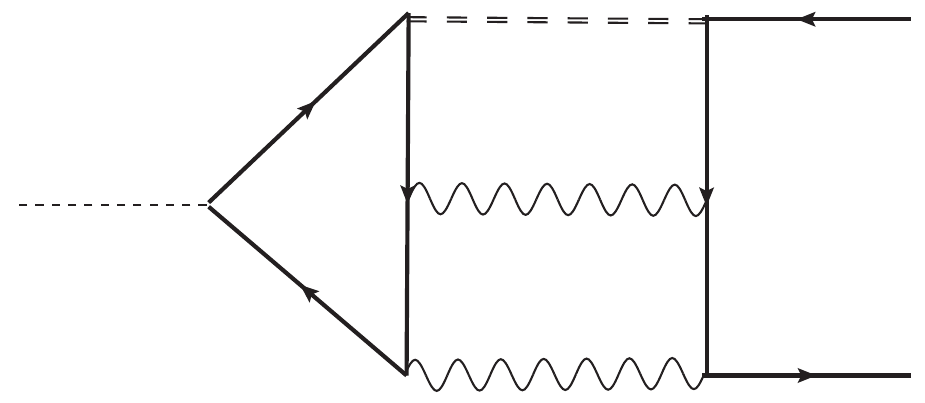}
\caption{
A representative diagram for the 3-loop corrections to the Yukawa coupling between an external Higgs boson (the external dashed line) and a pair of external massive fermions (the open solid lines with arrows) that contains a $\g5$-odd fermion loop represented by the solid circle with arrows.
The wavy lines represent gauge-boson propagators and the double-dashed line is either a gauge-boson or a would-be Goldstone propagator.
}
\label{fig:HiggsYukawaVertex_CFtype3L}
\end{center}
\end{figure}

Regarding diagrams with open fermion lines, the 3-point-vertex Green functions with a pair of external fermions and one external boson have a non-negative superficial UV-degree 0.
According to the discussions at the end of subsection~\ref{sec:epsion-MBA}, we are not aware of any argument that can be used to prevent the 3-loop corrections with a $\g5$-odd fermion loop to the Yukawa interaction between a Higgs scalar and a pair of fermions (as illustrated in figure~\ref{fig:HiggsYukawaVertex_CFtype3L}) from developing an overall UV-divergence.
Note, in particular, that this 3-loop diagram does \textit{not} vanish even if the momentum insertion through the external Higgs boson vanishes.  
Furthermore, the combined effect of incorporating contributions with the double-dashed line in figure~\ref{fig:HiggsYukawaVertex_CFtype3L} being a EW gauge-boson propagator with mass $M_a$ in $R_{\xi}$-gauge~\cite{tHooft:1972qbu,Fujikawa:1972fe}, 
\begin{equation} \label{eq:RxigaugeVBpropagator}
\frac{i\,\big(-g^{\mu\nu}+\frac{p^{\mu}p^{\nu}}{p^2-\xi_a M_a^2}\,(1-\xi)\big)}{p^2-M_a^2+i\epsF} = 
\frac{i\, \big(-g^{\mu\nu} + \frac{p^{\mu}p^{\nu}}{M_a^2}\big)}{p^2-M_a^2+i\epsF}  
 \,-\, \frac{i\, \frac{p^{\mu}p^{\nu}}{M_a^2}}{p^2-\xi M^2_a+i\epsF}
\end{equation} 
and the corresponding would-be Goldstone propagator $1/(p^2-\xi_a M^2_a+i\epsF)$ (with appropriate Yukawa couplings proportional to fermion masses) is equivalent to simply taking just this massive EW gauge-boson propagator in unitary gauge (given by the first $\xi_a$-independent term in the r.h.s.~of \eqref{eq:RxigaugeVBpropagator}).
Clearly, the fermion-propagator mass power factors resulting from the contraction with the numerator of $p^{\mu}p^{\nu}/M_a^2$ are not sufficient to render the proper UV-degree to be negative. 
The computations made in ref.~\cite{Chetyrkin:2012rz} contain 3-loop diagrams of the same structure as in figure~\ref{fig:HiggsYukawaVertex_CFtype3L}, with the double-dashed line there being a pseudo-scalar propagator (as a simplified SM Lagrangian without EW-gauge interactions was used in ref.~\cite{Chetyrkin:2012rz}, sufficient to capture the dominant contributions of interest);
the explicit results for these diagrams showed that they do contain $1/\epsilon$ UV poles.
The applicability of the procedure described in subsection~\ref{sec:pres_g5vertex} to the 3-loop diagram in figure~\ref{fig:HiggsYukawaVertex_CFtype3L} is not yet explicitly established (up to the finite $\epsilon^0$-order).

Based on the survey carried out above, all multi-boson diagrams with $\g5$-odd fermion loops up to 3-loop order in SM 
do not contain overall UV divergences. 
We call the SM diagrams where $\g5$-odd fermion loops appear but not inside any overall-divergent subgraph as Case-$A$ diagrams. 
SM loop diagrams with $\g5$-odd fermion loops inside an overall-divergent subgraphs will be classified as Case-$B$ diagrams, to be discussed in subsection~\ref{sec:caseB}.

\subsubsection{Treatment of the Levi-Civita tensors}
\label{sec:LeviCiviaTensor}

It is known that the (pseudo) Levi-Civita tensor $\epsilon^{\mu\nu\rho\sigma}$ can only be mathematically defined consistently in 4 dimensions, and mathematical inconsistency appears once one insists on the commutation in the contraction ordering for a product of multiple of them with an indefinite dimension D~\cite{Breitenlohner:1977hr,Siegel:1980qs}, due to the lack of the 4-dimensional Schouten identity.
Of course, one may consider using the standard Levi-Civita tensor defined with all Lorentz indices in 4 dimensions in addition to the D-dimensional ones carried by loop momenta in DR, when manipulating a product of them (each associated with one $\g5$-odd fermion loop). 
This is certainly doable, and operations like this are routinely performed in the symbolic computations based on a non-anticommuting BMHV $\g5$~\cite{tHooft:1972tcz,Breitenlohner:1977hr,Breitenlohner:1975hg,Breitenlohner:1976te} with the aid of a consistent dimensional splitting, and similarly in the KKS/Kreimer prescription~\cite{Korner:1991sx,Kreimer:1993bh}.

On the other hand, it would be interesting to know whether an alternative treatment using one and the only D-dimensional spacetime metric tensor could equally work, which may be technically convenient in symbolic computations.
We believe that the answer to this question for all diagrams of Case-$A$, where $\g5$-odd fermion loops are not inside any overall-divergent 1PI graphs, is positive, as far as the final 4-dimensional results for physical quantities are concerned. 
The essential logic behind this is as follows: 
after subtracting all UV divergences, none of which involve $\g5$-odd traces or contractions of pairs of $\epsilon^{\mu\nu\rho\sigma}$ in our treatment, the renormalized result as a whole is finite and has a smooth 4-dimensional $\epsilon \rightarrow 0$ limit\footnote{The intermediate IR divergences in amplitudes are assumed to be either absent or regularized by non-dimensional regulators at the moment for the simplicity of the discussion; alternatively, the subject of this statement shall be promoted to the expressions for IR-safe physical quantities in terms of squared amplitudes as a whole. See subsection~\ref{sec:g5trace_squaredamps} for more related discussion.};
consequently, $\epsilon^{\mu\nu\rho\sigma}$ appears in Case-A diagrams either explicitly as a single overall factor, or through contracted pairs $\epsilon^{\mu\nu\rho\sigma} \, \epsilon^{\mu'\nu'\rho'\sigma'}$ (each from one $\g5$-odd fermion loop), all multiplied onto finite quantities; 
these objects are thus allowed to be treated with $\epsilon$-suppressed spurious pieces included, for instance with Lorentz indices of the spacetime metric tensors from the contraction $\epsilon^{\mu\nu\rho\sigma} \, \epsilon^{\mu'\nu'\rho'\sigma'}$ set in D-dimensions,  
which will drop in the 4-dimensional limit provided that they are added consistently in every bare singular pieces.  
This can be easily achieved for Case-A diagrams as explained above.

For the sake of reader's convenience, below we recapitulate and streamline our treatment of $\epsilon^{\mu\nu\rho\sigma}$ (previously outlined in the appendix of~\cite{Chen:2023lus}).
Clearly there is one $\epsilon^{\mu\nu\rho\sigma}$ resulting from each $\g5$-odd trace associated with a closed fermion chain, manipulated according to the anchor prescription described in subsection~\ref{sec:pres_g5vertex} and implemented in \gfanchor to be introduced in subsection~\ref{sec:tech}.
The pair of $\epsilon^{\mu\nu\rho\sigma}$ to be contracted must come from $\g5$-odd traces associated with two \textit{different} fermion loops and/or external polarization-state projectors. 
We then proceed as follows.
\begin{itemize}
\item 
Denote the contraction of a general product of multiple Levi-Civita tensors in our problem by 
\begin{equation}
\label{eq:LVTC}
\big[\epsilon_{i_1}\, \, \, \cdots\, \epsilon_{i_{N_i}}\big]\,  
\big[\epsilon_{e_1}\, \, \cdots\, \epsilon_{e_{N_e}}\big] 
\end{equation}
where Lorentz indices are suppressed for the sake of concise notations, and the $[~]$ wrapping around a product of Levi-Civita tensors indicates the contraction.
Two sets of subscripts are introduced:
those labeled by $e_k$ for $k = 1,\cdots, N_e$ are from external polarization-state projectors; 
those labeled by $i_k$ for $k = 1,\cdots, N_i$ are from different $\g5$-odd fermion loops.

To have definite unique expressions for intermediate bare loop amplitudes (that is before subtracting all divergences and subsequently taking the 4-dimensional limit),  
we proceed in the following way.
\begin{itemize} 
\item 
For $\big[\epsilon_{e_1}\, \, \cdots\, \epsilon_{e_{N_e}}\big]$ in eq.\eqref{eq:LVTC} which appears as an overall factor for the quantities in question, simply mark each of them and fix an arbitrary but definite contraction-order, which shall be adopted consistently in the calculation. 

\item 
For $\big[\epsilon_{i_1}\, \, \, \cdots\, \epsilon_{i_{N_i}}\big]$ in eq.\eqref{eq:LVTC}, take a symmetric average over all possible pairings. 
For instance in the case of $N_i = 4$, we define 
\begin{eqnarray*}
\big[\epsilon_1\, \epsilon_2\, \epsilon_3\, \epsilon_4\big] 
\equiv 
\frac{1}{3} 
\big( 
\big[ \epsilon_1\, \epsilon_2 \big] 
\big[ \epsilon_3\, \epsilon_4 \big]
\,+\, 
\big[ \epsilon_1\, \epsilon_3 \big] 
\big[ \epsilon_2\, \epsilon_4 \big]
\,+\, 
\big[ \epsilon_1\, \epsilon_4 \big] 
\big[ \epsilon_2\, \epsilon_3 \big]
\big)\,.
\end{eqnarray*}
The numbers of the possible pairings to be averaged over in $\big[\epsilon_{i_1}\, \, \, \cdots\, \epsilon_{i_{N_i}}\big]$ are 
\begin{equation*}
\begin{cases}
& \frac{N_i !}{(N_i/2)!\,2^{N_i/2}}  \quad \text{if $N_i$ is even}\,, \\
& N_i\,\frac{(N_i-1)!}{((N_i-1)/2)!\,2^{(N_i-1)/2}}  \quad \text{if $N_i$ is odd}\,. 
\end{cases}
\end{equation*}

\item 
The results derived respectively for $\big[\epsilon_{i_1}\, \, \, \cdots\, \epsilon_{i_{N_i}}\big]$ and $\big[\epsilon_{e_1}\, \, \, \cdots\, \epsilon_{e_{N_e}}\big]$ are then multiplied together as in eq.\eqref{eq:LVTC}.
\end{itemize}

\item 
The contraction between any single pair of Levi-Civita tensors involved above is done according to the following standard formula
\begin{eqnarray} \label{eq:LeviCivitaContRule}
\epsilon^{\mu\nu\rho\sigma} \epsilon^{\mu'\nu'\rho'\sigma'} 
= \mathrm{Det}\Big[g^{\alpha \alpha'} \Big]~, 
\text{~  with $\alpha \in \{\mu,\nu,\rho,\sigma\}$ and $\alpha' \in \{\mu',\nu',\rho',\sigma'\}$,} 
\end{eqnarray}
but with the resulting spacetime-metric tensor $g_{\mu\nu}$ set \textit{$D$-dimensional}. 
\end{itemize}

Note, however, the above manual symmetrization in the pairing of Levi-Civita tensors~\cite{Zerf:2019ynn,Chen:2023lus} does not really restore the commutativity, nor the associative law for a product of multiple Levi-Civita tensors with (indefinite) D-dimensional Lorentz indices.
It should be regarded as an optional convenient trick to spare one from the tedious work of explicitly bookkeeping a particular and consistent choice of the contraction ordering for bare amplitudes at different loop orders (which shall be adopted consistently among those to be combined together to get renormalized ones), applicable at least for Case-$A$ diagrams. 
It is worthy emphasizing that the validity of the above treatment of Levi-Civita tensors without requiring 4-dimensional Lorentz indices in the r.h.s.~of \eqref{eq:LeviCivitaContRule} is not solely due to the uniqueness of the expressions for bare loop amplitudes determined in this prescription, 
but primarily owes to the fact that any possible $\epsilon$-suppressed overall deviation in the r.h.s.~of \eqref{eq:LeviCivitaContRule} from the strictly 4-dimensional treatment drops in the 4-dimensional limit for Case-$A$ diagrams.

\subsection{Case-$B$: $\g5$-odd fermion loops inside an overall-divergent 1PI amplitude}
\label{sec:caseB}
 
\begin{figure}[htbp]
\begin{center}
\includegraphics[scale=0.48]{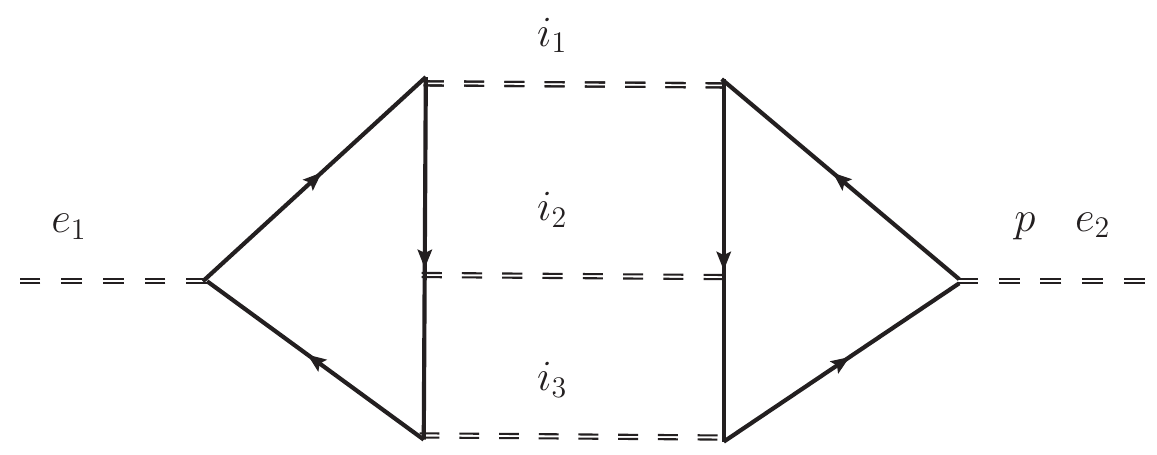}
\caption{
A prototype diagram for contributions to the self-energy functions of bosons 
with a pair $\g5$-odd fermion loops at 4-loop order. 
The solid boxes with arrow denote the $\g5$-odd fermion loops,
and the double-dashed lines (with labels $e_1,\, e_2$ for legs and $i_1,\, i_2,\, i_3$ for propagators) are either gauge bosons or scalars. The external momentum flowing through the diagram is denoted as $p$.
}
\label{fig:BosonSelfEnergy4L}
\end{center}
\end{figure}

As alluded in the discussions in subsection~\ref{sec:epsion-MBA}, there are Case-$B$ diagrams in SM, which contain $\g5$-odd fermion loops inside certain overall-divergent subgraphs.
In figure~\ref{fig:BosonSelfEnergy4L}, a representative diagram for the simplest examples at 4-loop order are illustrated, 
which are obtained by adding one more internal loop propagator between the two $\g5$-odd fermion loops in the 3-loop representative diagram in figure~\ref{fig:DoubleTriangle}.
The incorporation of one more leg to each of the $\g5$-odd fermion loop helps to alleviate the power-suppression in the large-loop momentum region arising from dependence of $\g5$-odd Dirac traces on external momentum $p$ or mass terms in the numerators of fermion propagators, which was crucial to ensure the UV-finiteness of the 3-loop counterparts.

The so-called non-naive 4-loop contributions to the wave-function renormalization of gauge bosons (hence renormalization of gauge couplings) in refs.~\cite{Zoller:2015tha,Bednyakov:2015ooa,Poole:2019txl,Davies:2019onf} correspond to the figure~\ref{fig:BosonSelfEnergy4L} with $e_1,\, e_2$ being gauge bosons and $i_1,\, i_2,\, i_3$ being gauge bosons, would-be Goldstones and Higgs bosons. 
The comments there on these 4-loop contributions to the wave-function renormalization remain valid in the $R_{\xi}$-gauge~\cite{tHooft:1972qbu,Fujikawa:1972fe} 
when the fermions are taken massive: 
the configurations with two of the three propagators $i_1,\, i_2,\, i_3$ being scalar and pseudo-scalar bosons lead to overall-UV divergent contributions featuring an overall factor of power 4 in fermion's Yukawa couplings and/or masses (divided by the Higgs vaccum-expectation value).  
Going to the unitary gauge by taking $\xi \rightarrow \infty$ can, however, shuffle the overall-divergent contributions among different diagrams generated by different particle identifications for $i_1,\, i_2,\, i_3$ propagators (as indicated by  \eqref{eq:RxigaugeVBpropagator}).
This can be seen most clearly by contemplating how to reproduce the 4-loop Top-Yukawa effects on the QCD $\beta$-function of the strong coupling $\alpha_s$ determined in refs.~\cite{Zoller:2015tha,Bednyakov:2015ooa} but in terms of SM Feynman diagrams in unitary gauge.

\subsubsection{An example of the problematic diagrams}
\label{sec:pdexample}

Despite the presence of the 4-loop Case-$B$ contribution in SM, such as those in figure~\ref{fig:BosonSelfEnergy4L}, it might be premature to conclude the failure of the original Kreimer's prescription~\cite{Kreimer:1993bh} (where certain revisions were incorporated compared to the earlier KKS version~\cite{Korner:1991sx}) for these diagrams.
Indeed, ref.~\cite{Bednyakov:2015ooa} pointed out explicitly that the correct result~\cite{Poole:2019txl,Davies:2019onf} can follow from a reading prescription labeled "C" in ref.~\cite{Bednyakov:2015ooa}; 
and it is not completely inconceivable to refine the original prescription~\cite{Kreimer:1993bh} slightly regarding the choice of $\g5$ anchor points such that only these valid ones are allowed.
In fact, these valid $\g5$ anchor points for the 4-loop Case-$B$ diagrams in figure~\ref{fig:BosonSelfEnergy4L} are the natural choices if one ignores the overall UV-divergence of the whole 4-loop diagrams and limits the scope just to the subgraphs containing exactly one $\g5$-odd fermion loop.
In particular, the aforementioned valid choices are precisely the $\g5$ anchor points determined by simply applying the procedure of subsection~\ref{sec:pres_g5vertex} to these 4-loop Case-$B$ diagrams as if they were Case-$A$ diagrams by ignoring their overall UV-divergence.  
~\\

\begin{figure}[htbp]
\begin{center}
\includegraphics[scale=0.48]{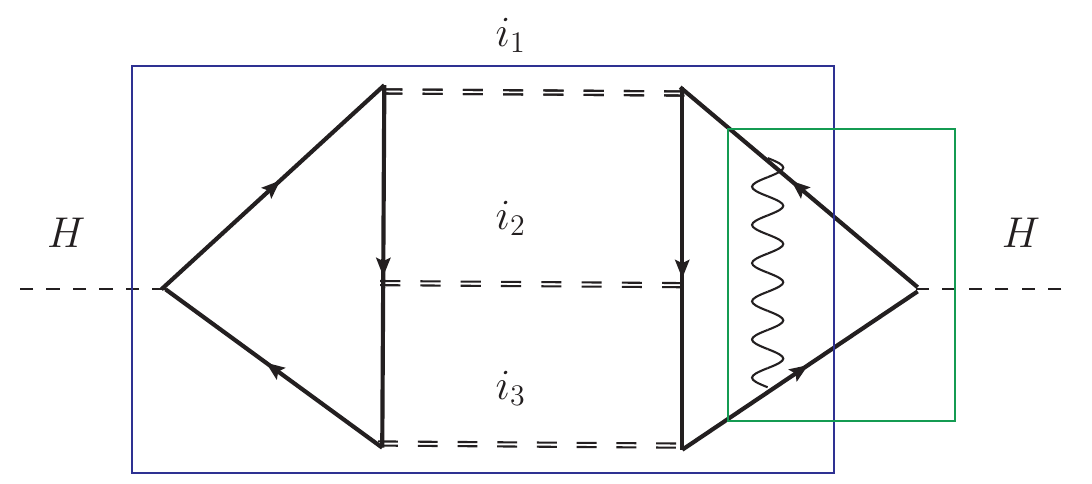}
\caption{
A representative 5-loop Case-$B$ diagram with a pair $\g5$-odd fermion loops contributing to the self-energy functions of the Higgs boson where there does not exist any qualified choice to anchor $\g5$. 
The solid circles with arrow denote the $\g5$-odd fermion loops,
and the double-dashed lines (with labels $i_1,\, i_2,\, i_3$ for propagators) are either gauge bosons or scalars.
The subgraph encircled in the blue box is essentially the UV-divergent Yukawa vertex~\ref{fig:HiggsYukawaVertex_CFtype3L} but with an additional virtual gauge-boson propagator, and the green box isolates another UV-divergent vertex subgraph that involves the same $\g5$-odd fermion loop.
}
\label{fig:HiggsSelfEnergy5L}
\end{center}
\end{figure}

However, combining the information discussed in the end of subsection~\ref{sec:epsion-MBA} and~\ref{sec:caseA}, in particular the  figure~\ref{fig:HiggsYukawaVertex_CFtype3L} is overall-divergent, we can compose the following example in SM shown in figure~\ref{fig:HiggsSelfEnergy5L}, 
where we believe that both the prescription~\cite{Kreimer:1993bh} and the procedure described in subsection~\ref{sec:pres_g5vertex}  can not be applied.   
The problem here is that there appears overlapping between the two sub-divergences, encircled in the blue and green boxes around the two Higgs Yukawa vertices respectively, which altogether cover all propagators on the same $\g5$-odd fermion loop.
\footnote{In loop diagrams free of the $\g5$ issue, overlapping UV divergences are common and pose no problem at all for UV renormalization in theories such as QCD. 
}
(The subgraph encircled in the blue box is essentially the UV-divergent Yukawa vertex~\ref{fig:HiggsYukawaVertex_CFtype3L} but with an additional virtual correction.)
Consequently, there does not exist any qualified choice to anchor $\g5$ on this fermion loop that fulfills the condition of being \textit{outside} all sub-divergences associated with vertex loop corrections as prescribed in subsection~\ref{sec:pres_g5vertex}.
~\\

The representative diagram we composed in figure~\ref{fig:HiggsSelfEnergy5L}, as well as similar ones obtained by distributing the additional virtual correction at various other places in the fermion loops, merely provides a particular counter-example where we think it is manifestly clear that the conditions for a valid $\g5$ anchor point as summarized in section~\ref{sec:pres_g5vertex} can not be fulfilled anyway.
But this does not necessarily imply that this anti-commuting-$\g5$-based treatment remains valid in SM all the way up to 4-loop orders, even though the results in ref.~\cite{Bednyakov:2015ooa} show that certain natural and valid $\g5$ anchor points do exist for the 4-loop Case-$B$ diagrams illustrated in figure~\ref{fig:BosonSelfEnergy4L}.  
Irrespective of the possible applicability to some 4-loop diagrams, it does not seem to be applicable to all orders in anomaly-free SM.

\subsubsection{Discussions on the multiplicative renormalizability to be checked}

As stated at the very beginning of this section~\ref{sec:pres}, since we do not have a solid proof for the validity of the proposed procedure to manipulate $\g5$ in DR, strictly speaking, it has not been established as a consistent scheme and thus maintains at present merely the status of a prescription.
Explicit sanity checks of whether it works as expected shall be performed. 
Obviously, the most crucial criteria is the multiplicative renormalizability of SM, following from the various defining WTs and STs (see, e.g.\,~refs.~\cite{Aoki:1982ed,Bohm:1986rj,Denner:1994xt,Kraus:1997bi,Bohm:2001yx,Denner:2019vbn,Belusca-Maito:2023wah}), shall now emerges \textit{automatically} from the so-prescribed computations. 
To this end, we should verify, for example, whether the local counter-terms in SM Lagrangian with the renormalization constants determined solely by a handful set of renormalization conditions 
are indeed sufficient to cancel all UV divergences in on-shell S-matrix elements in SM.

It is important to note, however, that having validated merely a special subset of WTs and/or STs, such as the so-called doubly-contracted ones involving just self-energy functions of EW bosons~\cite{Bohm:2001yx,Denner:2019vbn,Actis:2006ra,Actis:2006rb,Actis:2006rc}, is still far from certifying a prescription for treating $\g5$, because not much can be inferred from this regarding the consistency of the $\g5$ treatment in general. 
For example, there is not yet any $\g5$-odd fermion loop in the 1PI diagrams involved in these relations up to 2-loop order, which is clear from subsection~\ref{sec:caseA};
hence we do not expect any principal $\g5$-related difficulty to have these relations checked up to 2-loop order, such as explicitly done in ref.~\cite{Actis:2006ra,Actis:2006rb,Actis:2006rc} using simply a naive anticommuting $\g5$~\cite{Bardeen:1972vi,Chanowitz:1979zu,Gottlieb:1979ix,Abdelhafiz:1986jh}. 

We have shown in subsection~\ref{sec:caseA} that up to 3-loop order in SM fermion loops are either free of $\g5$ or belong to the Case-$A$, except for the non-trivial configuration illustrated in figure~\ref{fig:HiggsYukawaVertex_CFtype3L} where an explicit check is yet to be accomplished.
On the other hand, up to 3-loop order, the difference between our treatment of $\g5$ from the prescriptions~\cite{Korner:1991sx,Kreimer:1993bh} is not very essential, and mainly concerns the technical side and practical convenience, as well as possibly the recipe for those with intermediate IR divergences\footnote{To avoid confusion, note that even for diagrams below 4-loop orders, the bare expressions determined in our prescription are in general different from those using the prescriptions~\cite{Korner:1991sx,Kreimer:1993bh}, and shall not be compared naively at the bare level.} (to be discussed in the next subsection).
We would like to emphasize that explicit checks of the WTs and STs at 3-loop orders with $\g5$-odd fermion loops up to the $\epsilon^0$-order are not yet available with this kind of  $\g5$ treatment.
Furthermore, BMHV is known as a consistent regularization scheme that can be systematically carried out, in principle, to arbitrary perturbative order,
however, we do not have a proof to show that the prescription given in this manuscript differ from the BMHV just by some local, Hermitian counterterms, and hence could not demonstrate that it respects the unitarity postulate of QFT (e.g.~formulated in refs.~\cite{tHooft:1971akt,tHooft:1971qjg,tHooft:1972tcz,Becchi:1974xu,Becchi:1974md,Becchi:1975nq} and discussed in refs.~\cite{,Breitenlohner:1977hr,Breitenlohner:1975hg,Breitenlohner:1976te,Gnendiger:2017pys,Belusca-Maito:2023wah}).
Needless to say, this issue remains to be clarified in the future for the  application of the prescription at multi-loop orders.

\subsection{Traces for fermion chain closed in squared amplitudes}
\label{sec:g5trace_squaredamps}

The discussion of $\g5$ anchor points in the previous subsections are concerned with UV divergences in SM amplitudes with presumably \textit{fixed} external kinematics.
The possible IR divergences from loop integration in amplitudes as well as from phase-space integration of squared amplitudes are implicitly assumed to be either absent or regularized by certain non-dimensional regulators (such as auxiliary masses and/or off-shellness of external momenta) for simplicity. 
As different $\g5$ regularization prescriptions, as well as different $\g5$ anchor points, lead to in general bare loop amplitudes with different $\epsilon$-dependence, the intermediate IR divergences in the UV-renormalized amplitudes, if regularized also dimensionally using the same regulator $\epsilon$, may lead to spurious $\g5$-prescription-dependent pieces.
To ensure the cancellation of the possible spurious $\g5$-prescription dependence, any \textit{minimal} set of the IR-divergent contributions of which the IR divergences cancel in the sum (leading to a 4-dimensional finite IR-safe quantity), which will thus be referred to as \textit{IR-correlated} below, must not be determined completely independent of each other.
The following exposition can make it more clear to us how to proceed and why it may work.
~\\

As is well-known, it is the modulus-squared amplitudes, which may be conveniently represented in terms of the so-called cut-diagrams~\cite{Kinoshita:1962ur} in perturbation theory, that are directly related to physical observables such as the decay rates and cross sections. 
In the physical Minkowski region of the external kinematics, individual cut-diagrams may contain contributions from subsets of IR-degenerated quantum states, due to presence of massless particles in the interacting theory, leading to intermediate IR-divergences. 
According to Kinoshita-Lee-Nauenberg (KLN) theorem~\cite{Kinoshita:1962ur,Lee:1964is}, the cancellation of theses intermediate IR divergences shall be observed once they are collected together consistently from all individual cut-diagrams contributing to a properly defined IR-safe physical observable.
It is important to note that this holds independent of the details of the UV renormalization in the field theory, which are not even necessarily done provided all UV divergences are properly regularized~\cite{Kinoshita:1962ur}; 
moreover, it does not require, in principle, these IR divergences to exhibit factorization property~\cite{Kinoshita:1962ur,Lee:1964is} (although IR divergences do factorize 
to certain extent which helps to simplify the matter a lot in practice). 
Therefore, the condition to maintain the KLN cancellation mechanism of intermediate IR divergences for a $\g5$ prescription is quite different from those UV-related constraints described in subsection~\ref{sec:pres_g5vertex}: 
in principle, it is \textit{not} really necessary to pull $\g5$ outside any maximal 1PI vertex correction with an open fermion line as far as only the IR-divergence cancellation is concerned; 
in other words, $\g5$ could have been simply anchored where it was introduced by the Feynman rules in all IR-correlated contributions  
an option disfavored unfortunately merely because of the issue with a symmetry-preserving UV renormalization.

Therefore, according to our understanding, the key to ensure the KLN cancellation in a $\g5$ prescription lies in the compatibility of the definition of the $\g5$-odd Dirac traces with Cutkosky's cutting rule~\cite{Cutkosky:1960sp}, 
in other words, the diagrammatic-level Cutkosky's cutting rule shall remain applicable in the properly-devised $\g5$ prescription. 
To this end, we provide the following recipe. 
The cut-diagram representations for the IR-correlated squared amplitudes contributing 
at the given perturbative order share the same prototype uncut loop-diagrams, where the only difference is merely which particular set of $n$ Feynman propagators are put on-shell (i.e.~cut open) according to the following on-shell cutting rule~\cite{Cutkosky:1960sp},
\begin{equation} \label{eq:onshellcuts}
\frac{i}{p_j^2 - m_j^2 + i\epsF}\Big|_{\text{on-shell}} \,\rightarrow\, 2\pi\delta(p_j^2 - m_j^2)\Theta(p_j^0) \quad \text{for $j=1,2,\cdots, n$,}   
\end{equation}
with $\sum_{j=1}^{n} p_j$ equal to the total momentum flow through this cut;  
the Dirac trace associated with a $\g5$-odd closed fermion chain in this uncut loop-diagram shall be defined with the same definite $\g5$ anchor points, according to subsection~\ref{sec:pres_g5vertex}, \textit{without} considering, and hence irrespective of, the particular on-shell cuts on this fermion loop. 
(Note that a Case-$B$ diagram with on-shell cuts may be turned into a Case-$A$ diagram, due to elimination of the overall UV divergence by the on-shell cut, such as the 4-loop contributions of the type in figure~\ref{fig:BosonSelfEnergy4L} for the self-energy functions of gauge bosons.)

\begin{figure}[htbp]
\begin{center}
\includegraphics[scale=0.55]{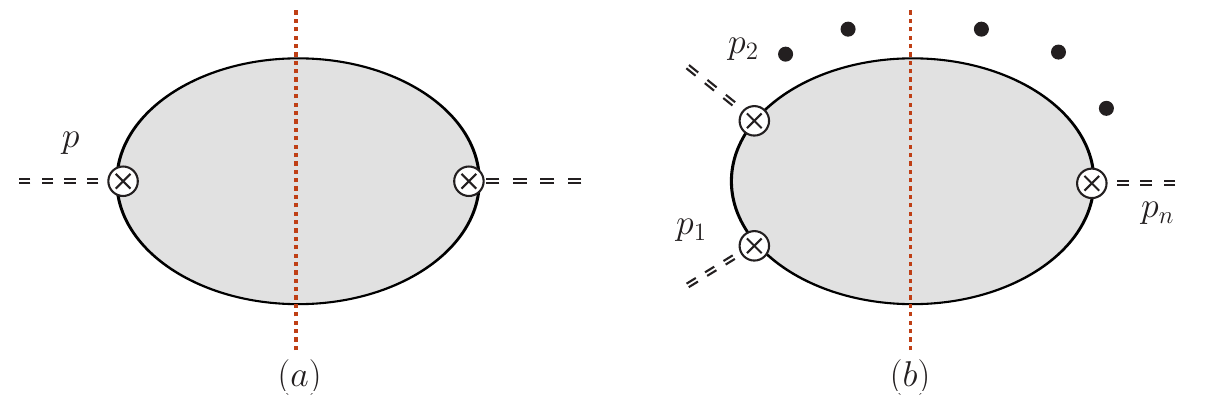}
\caption{
The illustration of closed fermion chains in cut-diagrams with possible on-shell cuts through them, represented by the red dotted lines.
The solid circles denote the $\g5$-odd closed fermion chain with shades representing possible virtual 
corrections.  
The double-dashed lines represent boson propagators external to the $\g5$-odd closed fermion chain in question. 
The category with exactly 2 external double-dashed lines on the target $\g5$-odd closed fermion chain is labeled as (a), where there is no any other $\g5$-odd fermion loops in the shades.
The block dots in (b) denote possible additional external double-dashed lines not drawn explicitly. 
}
\label{fig:CutDiagrams}
\end{center}
\end{figure}

The only exceptional configuration to the above general recipe is a 2-point subgraph with one $\g5$-odd fermion loop, which vanishes algebraically (due to the fully anti-symmetric Levi-Civita tensor) unless being cut open, such as illustrated in figure~\ref{fig:CutDiagrams}-(a), with \textit{non-inclusive} 
phase-space integration.\footnote{In fact, had an uncut 2-point $\g5$-odd fermion loop not been vanishing, we would be in trouble to apply our procedure to identify a $\g5$ anchor point for it: the MIOFV graphs associated with its two external legs overlap completely with each other, and thus there is no qualified $\g5$ anchor point.}  
On the other hand, since any non-vanishing contribution of this class must always have this loop cut open, 
this feature offers us a feasible alternative just for this exceptional case: 
all IR-correlated cut-diagrams of this special class share the same \textit{open} fermion line, and its two external on-shell spinors may be taken as the $\g5$ anchor points. 
Apparently this fully complies with the well-known treatment of the open-fermion chain in amplitudes as described in subsection~\ref{sec:pres_g5vertex}.

According to the discussions above, our recipe for determining the $\g5$ anchor points for closed fermion chains in IR-correlated squared amplitudes or cut-diagrams 
can be summarized as follows.
\begin{itemize}
\item n(>2)-point $\g5$-odd closed fermion chain in cut-diagrams, e.g.~figure~\ref{fig:CutDiagrams}-(b):~\\ 
The determination of the $\g5$ anchor points in this case shall be made for the target fermion loop without making any reference to, hence irrespective of, the possible on-shell cuts passing through it.
In short, the $\g5$ anchor points for a n(>2)-point $\g5$-odd closed fermion chain shall be determined as if it is not cut open, according to the procedure described in subsection~\ref{sec:pres_g5vertex}.

\item 2-point $\g5$-odd closed fermion chain in cut-diagrams, e.g.~figure~\ref{fig:CutDiagrams}-(a):~\\
 
It always identically vanishes without being cut open.
The two external on-shell spinors of the \textit{open} fermion line common among IR-correlated non-vanishing cut-diagrams of this special class can be taken as the $\g5$ anchor points.

We note that in this special case, the following replacement rule 
for the $\g5$ anchored behind an out-going on-shell fermion propagator (with momentum $p$ and mass $m$) can be applied, 
and may lead to Dirac traces with less powers of Dirac-$\gamma$ matrices compared to a direct application of \eqref{eq:gamma5}:
\begin{eqnarray}\label{eq:osIOspinors}
\frac{i}{\slashed{p} - m + i\epsF}\Big|_{\text{on-shell}} \, \gamma_5 
&=&  
2\pi\delta(p^2 - m^2)\Theta(p^0)\, 
\Big(\slashed{p} + m \Big)\, \gamma_5 \nonumber\\
&\rightarrow& 
2\pi\delta(p^2 - m^2)\Theta(p^0)\, 
\Big(
-\frac{i}{3!} \epsilon^{\mu\nu\rho\sigma} \gamma_{\nu} \gamma_{\rho} \gamma_{\sigma} p_{\mu}  
- \frac{i\, m}{4!}\epsilon^{\mu\nu\rho\sigma}\gamma_{\mu}\gamma_{\nu}\gamma_{\rho}\gamma_{\sigma}
\Big)\,,\nonumber\\
\end{eqnarray}
with $\epsilon^{\mu\nu\rho\sigma}$ treated as described in subsection~\ref{sec:LeviCiviaTensor}.
(Similar replacements were employed in ref.~\cite{Chen:2019wyb}, albeit in a different context.)
\end{itemize}
~\\

\begin{figure}[htbp]
\begin{center}
\includegraphics[scale=0.48]{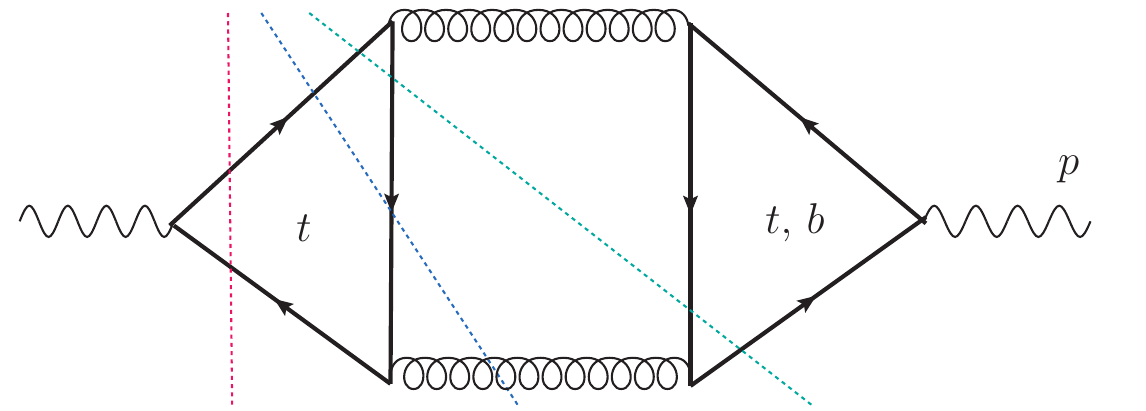}
\caption{
A representative of the cut-diagrams for the so-called triangle (interference) type QCD corrections at order $\alpha_s^2$~\cite{Catani:1999nf} to the $t\bar{t}$ production from a virtual Z boson (produced e.g.~in $e^+e^-$ collisions). 
The solid triangles with arrow denote the $\g5$-odd fermion loops,
and the external wavy lines are virtual Z bosons and the circular lines represent the gluons. 
The external momentum flowing through the diagram is denoted as $p$.
The red, blue and green dotted lines denote the 2-, 3- and 4-cut respectively, all passing necessarily through a pair of on-shell $t\bar{t}$ states by the definition of the $t\bar{t}$ production.
(The 4-cuts through four tops are excluded below four-top threshold.)
Other possible on-shell cuts of this uncut 3-loop diagram are not shown explicitly.
}
\label{fig:DoubleTriangleCuts}
\end{center}
\end{figure}

Now let us look at a concrete 3-loop example illustrated in figure~\ref{fig:DoubleTriangleCuts}, that is also very illuminating regarding the following point: it is \textit{not} always fine to shift blindly any $\g5$ on an amplitude-level open-fermion chain to external on-shell spinors without paying attention to the configurations in the cut-diagrams for squared-amplitudes. 
Figure~\ref{fig:DoubleTriangleCuts} shows some representative cut-diagrams for a special class of QCD corrections at order $\alpha_s^2$ to the top-quark pair $t\bar{t}$ production from a virtual Z boson produced e.g.~in $e^+e^-$ collisions, which are characterized by having both Z bosons coupled to two different fermion loops in the cut-diagrams. 
Owing to the Furry's theorem, the contribution to the inclusive $t\bar{t}$ production cross-section from this class of triangle-interference diagrams must have both Z-vertices being completely axial;
hence both fermion triangles in figure~\ref{fig:DoubleTriangleCuts} have an odd number of $\g5$.
In the approximation where only the top quark is kept massive among all quarks, one just needs to include the quarks of the third generation, i.e.~the top and (massless) bottom quark in the fermion triangles. 
The 2-, 3- and 4-cut in question all necessarily pass through a pair of on-shell $t\bar{t}$ states in figure~\ref{fig:DoubleTriangleCuts}, according to the definition of the $t\bar{t}$ production process.

It is clear that the triangle $\g5$-odd fermion loops in the Case-$A$ figure~\ref{fig:DoubleTriangleCuts} belong to the case of figure~\ref{fig:CutDiagrams}-(b).
Consequently, the discussion above tells us that the proper $\g5$ anchor points shall be the axial Z-coupling vertex, irrespective of the on-shell cuts on these fermion loops.
It is straightforward to see that this treatment will lead to the same finite 4-dimensional result as in the BMHV scheme. 
However, if shifting any $\g5$ on an amplitude-level open-fermion chain anticommutatively to external on-shell spinors was applied blindly as a doctrine, then $\g5$ in the 4-cut diagrams (being the product of two tree-level amplitudes) would be anchored at the on-shell cut propagators.
However, if shifting any $\g5$ on an amplitude-level open-fermion chain anticommutatively to external on-shell spinors were applied blindly as a doctrine, then $\g5$ on the triangle subgraphs with cuts would be anchored at the on-shell cut propagators.
We conducted an explicit check showing that below four-top threshold with the b-quark taken massless, 
this naive treatment led to a different, albeit still finite, value for the total sum of cut-diagrams in figure~\ref{fig:DoubleTriangleCuts}.
~\\

It may be appropriate to make the following comment at this point.
If some intermediate IR-subtraction terms are introduced separately for both virtual- and real-type contributions to a given physical observable in their respective partonic-level phase-spaces, then the pieces qualified as IR-correlated are further sub-divided; 
consequently, the above consistency condition in determining $\g5$ anchor points needs to be imposed just among the truly IR-correlated bare pieces whose combination lead to 4-dimensional finite remainders 
in each individual partonic phase-space.
By the virtue of the factorization of IR divergences, this refined version of consistency condition is usually easier to be achieved (among the singular pieces sharing exactly the same external states in each partonic phase-space), as compared to the original treatment without introducing intermediate IR-subtraction terms.

\subsection{Anchor $\g5$ automatically with \gfanchor}
\label{sec:tech} 

The procedure described in subsection~\ref{sec:pres_g5vertex} to anchor $\g5$ in Dirac traces in DR associated with closed fermion chains in SM Feynman diagrams, taking into account the insight in subsection~\ref{sec:g5trace_squaredamps} for cut-diagrams, is implemented in the program \gfanchor. 
The source code written in Mathematica can be found at%
\begin{center}
\url{https://gitlab.com/LongChenSDU/g5anchor}    
\end{center}

More specifically, \gfanchor finds and returns the $\g5$ anchor points for $\g5$-odd fermion loops in an input Feynman diagram whose graph structure and momentum flow are provided in a specific format (See below for more descriptions).
This information can then be passed to computer programs such as FORM~\cite{Vermaseren:2000nd}, FeynCalc~\cite{Mertig:1990an,Shtabovenko:2016sxi,Shtabovenko:2023idz}, PackagX~\cite{Patel:2015tea}, CalcLoop~\cite{calcloop} etc.~that have efficient implementation of the standard (cyclic) Dirac algebra, to obtain explicit expressions for $\g5$-odd Dirac traces in DR.

In the current implementation, the input diagrams are assumed to be effectively Case-$A$ diagrams defined in subsection~\ref{sec:caseA}, which in the ideal scenario may be the only type of diagrams that ever need non-trivial manipulations of $\g5$. 
Clear from our discussions above, this condition is at least satisfied for all cut-diagrams in the full SM up to 3-loop order (without introducing any non-dimensional regulator);
it is fulfilled, furthermore, even at higher loop orders for the QCD$\otimes$QED corrections to the scattering processes with fixed numbers of EW bosons.
The current implementation is partly checked using limited examples, up to 3-loop orders in QCD, available to the author so far.
~\\

\begin{figure}[htbp]
\begin{center}
\includegraphics[scale=0.9]{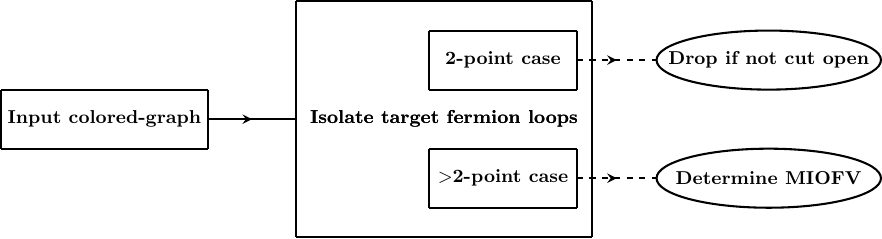}
\caption{
A simple illustration of the flowchart of the main function \textbf{AnchorG5} of \gfanchor 
}
\label{fig:flowchart}
\end{center}
\end{figure}

For a quick overview of what the package \gfanchor is used for and how to use the functions defined therein, one may start with the usage description of the main function \textbf{AnchorG5} and the variable  \textbf{Max1PIopenVFFsIOlegs} after loading the package in a Mathematica session.
The flowchart of the main function \textbf{AnchorG5} of \gfanchor is exposed in figure~\ref{fig:flowchart}.
Here we give a brief explanation of the workflow, and more technical information can be found in the detailed descriptions of the routines provided in the package.

The most important input for the function \textbf{AnchorG5} is a symbolic expression for a Feynman diagram $\mathrm{\mathbf{G}}$ containing $\g5$-odd fermion loops of interest, currently assumed to be Case-$A$, 
and the information on the graph connectivity and particle flow of $\mathrm{\mathbf{G}}$ shall be provided in a list of structured elements.
More specifically, the input $\mathrm{\mathbf{G}}$ is represented by a list of extended edges, each being one unique ordered list of four elements encoding the following information for one propagator:
the first element is a propagator label (which can be a number) unique in $\mathrm{\mathbf{G}}$;
the second element is a pair of directed vertex or node labels for the propagator;
the third element encodes the information on the particle identity of the propagator (such as electron or Z boson); 
the last element is the momentum along this propagator determined along the edge direction indicated by the aforementioned second element. 
Such a list representation for a propagator is called "colored-edge" in \gfanchor, and accordingly a list of them defines the "colored-graph" representation for the input Feynman diagram. 
See the description of the usage of function \textbf{ExtractColoredgraph} for more detail on this format.

Starting with this information of $\mathrm{\mathbf{G}}$, \textbf{AnchorG5} identifies the set of external momenta $E_Q$ of the target $\g5$-odd fermion loop $F_G$ by searching for the cut through a minimal number of boson propagators of the Case-$A$ diagram $\mathrm{\mathbf{G}}$ to isolate $F_G$ into a 1PI-diagram $G$ that contains no other $\g5$-odd fermion loops (except for fully nested ones in the current implementation), and has each of its external momenta equal to the difference between the momenta of certain pair of fermion propagators of $F_G$.  
The set $E_Q$ of the external momenta of $G$ then serves as the input for the next step, to determine the $\g5$ anchor points through identifying all (non-overlapping) MIOFV $\oVFF$ related to $F_G$; 
this information will be used to define the corresponding $\g5$-odd Dirac traces associated with $F_c$ as described in the end of subsection~\ref{sec:pres_g5vertex} and~\ref{sec:g5trace_squaredamps}.
More specifically, if there is just one independent momentum in $E_Q$, it is a 2-point $\g5$-odd fermion loop which will be kept only if cut open: the two external on-shell spinors of the \textit{open} fermion line are then taken as the $\g5$ anchor points in this case.
Otherwise, it is a  $\g5$-odd fermion loop with at least 3 external legs; the $\g5$ anchor points shall then be determined irrespective of the possible on-shell cuts going through this fermion loop.
~\\

There are also a few utility routines implemented in \gfanchor, such as the function \textbf{ManipulateG5inChain} and \textbf{DrawColoredGraphsIOBs}, which may be useful.
We have provided detailed explanation regarding their usage in \gfanchor.
Future updates of \gfanchor will include the functionality to determine the $\g5$ anchor points for closed fermion chains in the squared amplitudes 
directly based on the input colored-graph representations of the amplitudes in question.
The \gfanchor is primarily intended as a proof-of-principle implementation of the ideas discussed in this work, and any adaption or modification of the procedure to better address the specific problems at hand is welcomed.

\section{Conclusion}
\label{sec:conc}

Based on a recent revision of the works by Kreimer, Gottlieb and Donohue, we have reformulated the treatment of Dirac traces with $\g5$ in Dimensional Regularization, maintaining $\g5$'s anticommutativity only formally to certain extent, solely in terms of the usual notions such as the standard Dirac algebra apart from the non-4-dimensional treatment of Levi-Civita tensors;
the prescription is implemented in the procedure \gfanchor presented in this work.
We hope that this reformulation helps to make it more clear why we would expect the prescription to work in the cases where it does, and where failure might be expected instead.

Certain limitations and modifications of the KKS~\cite{Korner:1991sx} and/or the Kreimer~\cite{Kreimer:1993bh} scheme are discussed in light of ref.~\cite{Gottlieb:1979ix}. 
In addition to the unexpected failure when applied to amplitudes with external axial anomaly~\cite{Chen:2023lus}, the reason why they may not be applicable in SM beyond three or four loops in general is illustrated with examples: there are Case-$B$ diagrams with Higgs Yukawa vertices where certain conditions previously argued in refs.~\cite{Korner:1991sx,Kreimer:1993bh} no longer hold, and no valid $\g5$ anchor points can be found on dimensionally-regularized $\g5$-odd fermion loops in general (despite the accidental applicability in some four-loop gauge-boson self-energy diagrams). 
On the other hand, the prescription is expected to work for Case-$A$ diagrams (defined in subsection~\ref{sec:caseA}), in the sense of no need to manually incorporate $\g5$-related gauge-symmetry-restoration terms. 
It is important to note that explicit establishment of the Ward/Slavnov-Taylor identities involving Feynman diagrams with $\g5$-odd fermion loops are not yet available with this kind of $\g5$ treatment at 3-loop orders.
Whether the prescription differs from the well-established BMHV scheme just by some local, Hermitian counterterms remains to be clarified as well, which we hope to return in the near future.

From the procedure \gfanchor results the defining expression for the $\g5$-odd Dirac trace in Dimensional Regularization associated with a closed fermion chain in amplitudes, or more generally squared amplitudes, which we suggest to use in practical perturbative calculations in the Standard Model, at least up to two-loop order.
We note that this procedure \gfanchor itself shall be viewed as an implicit definition for a pseudo-anticommuting $\g5$ in this prescription with standard Dirac traces, which is neither completely non-anticommuting as in BMHV scheme nor truly anticommuting as in the 4-dimensional Dirac algebra.
The procedure formulated in this work, encoded in \gfanchor, shall be helpful, at least, in postponing the manual inclusion of $\g5$-related symmetry-restoration counter-terms into the SM Lagrangian, ideally, beyond three-loop order. 
For SM in absence of Yukawa couplings to Higgs field, especially for the QCD$\otimes$QED corrections to on-shell Green correlation functions involving local-composite operators with $\g5$, both $\g5$-related symmetry-restoration counter-terms and the dimensional-splitting needed to use 4-dimensional Levi-Civita tensors maybe be conveniently avoided at even higher loop orders with this procedure. 
As one of the interesting future applications of \gfanchor, one may extend to effective field theories with chiral local-composite operators in the effective Lagrangian, in particular the so-called Standard-Model Effective Field Theory, to one-loop order where we do not expect any essential difficulty.

While this work has clarified several issues for the good, it is essential to acknowledge that the validity of \gfanchor in SM up to three-loop order needs more explicit checks, which we plan to do in the near future.

\section*{Acknowledgments}

The author is grateful to Yan-Qing Ma, M.~Czakon, W.~Bernreuther, Yu Jia, Jian-Xiong Wang, Zong-Guo Si, Shi-Yuan Li, Jian Wang, Hai-Tao Li for their valuable discussions and/or feedbacks on the manuscript. 
Additionally, special thanks are extended to Zhe~Li, Xiang~Chen, Xin~Guan, Marco~Niggetiedt, Guo-Xing Wang, Bin Zhou, Peng-Cheng~Lu for various communications and/or for reading the manuscript.
This work was supported by the Natural Science Foundation of China under contract No.~12205171, No.~12235008, No.~12321005, and by the Taishan Scholar Foundation of Shandong province (tsqn202312052) and 2024HWYQ-005.

\bibliography{g5anchor} 

\providecommand{\href}[2]{#2}\begingroup\raggedright\begin{thebibliography}{100}

\bibitem{tHooft:1971akt}
G.~'t~Hooft, {\it {Renormalization of Massless Yang-Mills Fields}},
  \href{http://dx.doi.org/10.1016/0550-3213(71)90395-6}{{\em Nucl. Phys. B}
  {\bfseries 33} (1971) 173--199}.

\bibitem{tHooft:1971qjg}
G.~'t~Hooft, {\it {Renormalizable Lagrangians for Massive Yang-Mills Fields}},
  \href{http://dx.doi.org/10.1016/0550-3213(71)90139-8}{{\em Nucl. Phys. B}
  {\bfseries 35} (1971) 167--188}.

\bibitem{tHooft:1972tcz}
G.~'t~Hooft and M.~J.~G. Veltman, {\it {Regularization and Renormalization of
  Gauge Fields}},
\href{http://dx.doi.org/10.1016/0550-3213(72)90279-9}{{\em Nucl. Phys.}
  {\bfseries B44} (1972) 189--213}.

\bibitem{tHooft:1972qbu}
G.~'t~Hooft and M.~J.~G. Veltman, {\it {Combinatorics of gauge fields}},
  \href{http://dx.doi.org/10.1016/S0550-3213(72)80021-X}{{\em Nucl. Phys. B}
  {\bfseries 50} (1972) 318--353}.

\bibitem{Lee:1972fj}
B.~W. Lee and J.~Zinn-Justin, {\it {Spontaneously Broken Gauge Symmetries Part
  1: Preliminaries}},  \href{http://dx.doi.org/10.1103/PhysRevD.5.3121}{{\em
  Phys. Rev. D} {\bfseries 5} (1972) 3121--3137}.

\bibitem{Lee:1972ocr}
B.~W. Lee and J.~Zinn-Justin, {\it {Spontaneously Broken Gauge Symmetries Part
  2: Perturbation Theory and Renormalization}},
  \href{http://dx.doi.org/10.1103/PhysRevD.5.3137}{{\em Phys. Rev. D}
  {\bfseries 5} (1972) 3137--3155}. [Erratum: Phys.Rev.D 8, 4654 (1973)].

\bibitem{Lee:1972yfa}
B.~W. Lee and J.~Zinn-Justin, {\it {Spontaneously Broken Gauge Symmetries Part
  3: Equivalence}},  \href{http://dx.doi.org/10.1103/PhysRevD.5.3155}{{\em
  Phys. Rev. D} {\bfseries 5} (1972) 3155--3160}.

\bibitem{Weinberg:1971fb}
S.~Weinberg, {\it {Physical Processes in a Convergent Theory of the Weak and
  Electromagnetic Interactions}},
  \href{http://dx.doi.org/10.1103/PhysRevLett.27.1688}{{\em Phys. Rev. Lett.}
  {\bfseries 27} (1971) 1688--1691}.

\bibitem{Fujikawa:1972fe}
K.~Fujikawa, B.~W. Lee, and A.~I. Sanda, {\it {Generalized Renormalizable Gauge
  Formulation of Spontaneously Broken Gauge Theories}},
  \href{http://dx.doi.org/10.1103/PhysRevD.6.2923}{{\em Phys. Rev. D}
  {\bfseries 6} (1972) 2923--2943}.

\bibitem{Weinberg:1973ew}
S.~Weinberg, {\it {General Theory of Broken Local Symmetries}},
  \href{http://dx.doi.org/10.1103/PhysRevD.7.1068}{{\em Phys. Rev. D}
  {\bfseries 7} (1973) 1068--1082}.

\bibitem{Weinberg:1973ua}
S.~Weinberg, {\it {Perturbative Calculations of Symmetry Breaking}},
  \href{http://dx.doi.org/10.1103/PhysRevD.7.2887}{{\em Phys. Rev. D}
  {\bfseries 7} (1973) 2887--2910}.

\bibitem{Becchi:1974xu}
C.~Becchi, A.~Rouet, and R.~Stora, {\it {The Abelian Higgs-Kibble Model.
  Unitarity of the S Operator}},
  \href{http://dx.doi.org/10.1016/0370-2693(74)90058-6}{{\em Phys. Lett. B}
  {\bfseries 52} (1974) 344--346}.

\bibitem{Becchi:1974md}
C.~Becchi, A.~Rouet, and R.~Stora, {\it {Renormalization of the Abelian
  Higgs-Kibble Model}},
\href{http://dx.doi.org/10.1007/BF01614158}{{\em Commun. Math. Phys.}
  {\bfseries 42} (1975) 127--162}.

\bibitem{Becchi:1975nq}
C.~Becchi, A.~Rouet, and R.~Stora, {\it {Renormalization of Gauge Theories}},
\href{http://dx.doi.org/10.1016/0003-4916(76)90156-1}{{\em Annals Phys.}
  {\bfseries 98} (1976) 287--321}.

\bibitem{Aoki:1982ed}
K.~I. Aoki, Z.~Hioki, M.~Konuma, R.~Kawabe, and T.~Muta, {\it {Electroweak
  Theory. Framework of On-Shell Renormalization and Study of Higher Order
  Effects}},  \href{http://dx.doi.org/10.1143/PTPS.73.1}{{\em Prog. Theor.
  Phys. Suppl.} {\bfseries 73} (1982) 1--225}.

\bibitem{Bohm:1986rj}
M.~Bohm, H.~Spiesberger, and W.~Hollik, {\it {On the One Loop Renormalization
  of the Electroweak Standard Model and Its Application to Leptonic
  Processes}},  \href{http://dx.doi.org/10.1002/prop.19860341102}{{\em Fortsch.
  Phys.} {\bfseries 34} (1986) 687--751}.

\bibitem{Denner:1994xt}
A.~Denner, G.~Weiglein, and S.~Dittmaier, {\it {Application of the background
  field method to the electroweak standard model}},
  \href{http://dx.doi.org/10.1016/0550-3213(95)00037-S}{{\em Nucl. Phys. B}
  {\bfseries 440} (1995) 95--128},
  \href{http://arxiv.org/abs/hep-ph/9410338}{{\ttfamily arXiv:hep-ph/9410338}}.

\bibitem{Kraus:1997bi}
E.~Kraus, {\it {Renormalization of the Electroweak Standard Model to All
  Orders}},  \href{http://dx.doi.org/10.1006/aphy.1997.5746}{{\em Annals Phys.}
  {\bfseries 262} (1998) 155--259},
  \href{http://arxiv.org/abs/hep-th/9709154}{{\ttfamily arXiv:hep-th/9709154}}.

\bibitem{Grassi:1999nb}
P.~A. Grassi, {\it {Renormalization of nonsemisimple gauge models with the
  background field method}},
  \href{http://dx.doi.org/10.1016/S0550-3213(99)00457-5}{{\em Nucl. Phys. B}
  {\bfseries 560} (1999) 499--550},
  \href{http://arxiv.org/abs/hep-th/9908188}{{\ttfamily arXiv:hep-th/9908188}}.

\bibitem{Bohm:2001yx}
M.~Bohm, A.~Denner, and H.~Joos,
  \href{http://dx.doi.org/10.1007/978-3-322-80160-9}{{\em {Gauge theories of
  the strong and electroweak interaction}}}.
\newblock 2001.

\bibitem{Denner:2019vbn}
A.~Denner and S.~Dittmaier, {\it {Electroweak Radiative Corrections for
  Collider Physics}},
  \href{http://dx.doi.org/10.1016/j.physrep.2020.04.001}{{\em Phys. Rept.}
  {\bfseries 864} (2020) 1--163},
  \href{http://arxiv.org/abs/1912.06823}{{\ttfamily arXiv:1912.06823
  [hep-ph]}}.

\bibitem{Belusca-Maito:2023wah}
H.~B\'elusca-Ma\"\i{}to, A.~Ilakovac, P.~K\"uhler,
  M.~Ma\dj{}or-Bo\v{z}inovi\'c, D.~St\"ockinger, and M.~Wei\ss{}wange, {\it
  {Introduction to Renormalization Theory and Chiral Gauge Theories in
  Dimensional Regularization with Non-Anticommuting $\gamma_5$}},  {\em
  Symmetry} {\bfseries 15} (2023) 622,
  \href{http://arxiv.org/abs/2303.09120}{{\ttfamily arXiv:2303.09120
  [hep-ph]}}.

\bibitem{Freitas:2002ja}
A.~Freitas, W.~Hollik, W.~Walter, and G.~Weiglein, {\it {Electroweak two loop
  corrections to the $M_W-M_Z$ mass correlation in the standard model}},
  \href{http://dx.doi.org/10.1016/S0550-3213(02)00243-2}{{\em Nucl. Phys. B}
  {\bfseries 632} (2002) 189--218},
  \href{http://arxiv.org/abs/hep-ph/0202131}{{\ttfamily arXiv:hep-ph/0202131}}.
  [Erratum: Nucl.Phys.B 666, 305--307 (2003)].

\bibitem{Awramik:2002vu}
M.~Awramik, M.~Czakon, A.~Onishchenko, and O.~Veretin, {\it {Bosonic
  corrections to Delta r at the two loop level}},
  \href{http://dx.doi.org/10.1103/PhysRevD.68.053004}{{\em Phys. Rev. D}
  {\bfseries 68} (2003) 053004},
  \href{http://arxiv.org/abs/hep-ph/0209084}{{\ttfamily arXiv:hep-ph/0209084}}.

\bibitem{Actis:2006ra}
S.~Actis, A.~Ferroglia, M.~Passera, and G.~Passarino, {\it {Two-Loop
  Renormalization in the Standard Model. Part I: Prolegomena}},
  \href{http://dx.doi.org/10.1016/j.nuclphysb.2007.04.021}{{\em Nucl. Phys. B}
  {\bfseries 777} (2007) 1--34},
  \href{http://arxiv.org/abs/hep-ph/0612122}{{\ttfamily arXiv:hep-ph/0612122}}.

\bibitem{Actis:2006rb}
S.~Actis and G.~Passarino, {\it {Two-Loop Renormalization in the Standard Model
  Part II: Renormalization Procedures and Computational Techniques}},
  \href{http://dx.doi.org/10.1016/j.nuclphysb.2007.03.043}{{\em Nucl. Phys. B}
  {\bfseries 777} (2007) 35--99},
  \href{http://arxiv.org/abs/hep-ph/0612123}{{\ttfamily arXiv:hep-ph/0612123}}.

\bibitem{Actis:2006rc}
S.~Actis and G.~Passarino, {\it {Two-Loop Renormalization in the Standard Model
  Part III: Renormalization Equations and their Solutions}},
  \href{http://dx.doi.org/10.1016/j.nuclphysb.2007.04.027}{{\em Nucl. Phys. B}
  {\bfseries 777} (2007) 100--156},
  \href{http://arxiv.org/abs/hep-ph/0612124}{{\ttfamily arXiv:hep-ph/0612124}}.

\bibitem{Mihaila:2012fm}
L.~N. Mihaila, J.~Salomon, and M.~Steinhauser, {\it {Gauge Coupling Beta
  Functions in the Standard Model to Three Loops}},
  \href{http://dx.doi.org/10.1103/PhysRevLett.108.151602}{{\em Phys. Rev.
  Lett.} {\bfseries 108} (2012) 151602},
  \href{http://arxiv.org/abs/1201.5868}{{\ttfamily arXiv:1201.5868 [hep-ph]}}.

\bibitem{Mihaila:2012pz}
L.~N. Mihaila, J.~Salomon, and M.~Steinhauser, {\it {Renormalization constants
  and beta functions for the gauge couplings of the Standard Model to
  three-loop order}},  \href{http://dx.doi.org/10.1103/PhysRevD.86.096008}{{\em
  Phys. Rev. D} {\bfseries 86} (2012) 096008},
  \href{http://arxiv.org/abs/1208.3357}{{\ttfamily arXiv:1208.3357 [hep-ph]}}.

\bibitem{Bednyakov:2012rb}
A.~V. Bednyakov, A.~F. Pikelner, and V.~N. Velizhanin, {\it {Anomalous
  dimensions of gauge fields and gauge coupling beta-functions in the Standard
  Model at three loops}},
  \href{http://dx.doi.org/10.1007/JHEP01(2013)017}{{\em JHEP} {\bfseries 01}
  (2013) 017}, \href{http://arxiv.org/abs/1210.6873}{{\ttfamily arXiv:1210.6873
  [hep-ph]}}.

\bibitem{Martin:2015eia}
S.~P. Martin, {\it {Four-Loop Standard Model Effective Potential at Leading
  Order in QCD}},  \href{http://dx.doi.org/10.1103/PhysRevD.92.054029}{{\em
  Phys. Rev. D} {\bfseries 92} no.~5, (2015) 054029},
  \href{http://arxiv.org/abs/1508.00912}{{\ttfamily arXiv:1508.00912
  [hep-ph]}}.

\bibitem{Zoller:2015tha}
M.~F. Zoller, {\it {Top-Yukawa effects on the $\beta$-function of the strong
  coupling in the SM at four-loop level}},
  \href{http://dx.doi.org/10.1007/JHEP02(2016)095}{{\em JHEP} {\bfseries 02}
  (2016) 095}, \href{http://arxiv.org/abs/1508.03624}{{\ttfamily
  arXiv:1508.03624 [hep-ph]}}.

\bibitem{Chetyrkin:2016ruf}
K.~G. Chetyrkin and M.~F. Zoller, {\it {Leading QCD-induced four-loop
  contributions to the \ensuremath{\beta}-function of the Higgs self-coupling
  in the SM and vacuum stability}},
  \href{http://dx.doi.org/10.1007/JHEP06(2016)175}{{\em JHEP} {\bfseries 06}
  (2016) 175}, \href{http://arxiv.org/abs/1604.00853}{{\ttfamily
  arXiv:1604.00853 [hep-ph]}}.

\bibitem{Bednyakov:2015ooa}
A.~V. Bednyakov and A.~F. Pikelner, {\it {Four-loop strong coupling
  beta-function in the Standard Model}},
  \href{http://dx.doi.org/10.1016/j.physletb.2016.09.007}{{\em Phys. Lett. B}
  {\bfseries 762} (2016) 151--156},
  \href{http://arxiv.org/abs/1508.02680}{{\ttfamily arXiv:1508.02680
  [hep-ph]}}.

\bibitem{Davies:2019onf}
J.~Davies, F.~Herren, C.~Poole, M.~Steinhauser, and A.~E. Thomsen, {\it {Gauge
  Coupling $\beta$ Functions to Four-Loop Order in the Standard Model}},
  \href{http://dx.doi.org/10.1103/PhysRevLett.124.071803}{{\em Phys. Rev.
  Lett.} {\bfseries 124} no.~7, (2020) 071803},
  \href{http://arxiv.org/abs/1912.07624}{{\ttfamily arXiv:1912.07624
  [hep-ph]}}.

\bibitem{Baikov:2016tgj}
P.~Baikov, K.~Chetyrkin, and J.~K\"uhn, {\it {Five-Loop Running of the QCD
  coupling constant}},
  \href{http://dx.doi.org/10.1103/PhysRevLett.118.082002}{{\em Phys. Rev.
  Lett.} {\bfseries 118} no.~8, (2017) 082002},
  \href{http://arxiv.org/abs/1606.08659}{{\ttfamily arXiv:1606.08659
  [hep-ph]}}.

\bibitem{Herzog:2017ohr}
F.~Herzog, B.~Ruijl, T.~Ueda, J.~A.~M. Vermaseren, and A.~Vogt, {\it {The
  five-loop beta function of Yang-Mills theory with fermions}},
  \href{http://dx.doi.org/10.1007/JHEP02(2017)090}{{\em JHEP} {\bfseries 02}
  (2017) 090}, \href{http://arxiv.org/abs/1701.01404}{{\ttfamily
  arXiv:1701.01404 [hep-ph]}}.

\bibitem{Luthe:2017ttg}
T.~Luthe, A.~Maier, P.~Marquard, and Y.~Schroder, {\it {The five-loop Beta
  function for a general gauge group and anomalous dimensions beyond Feynman
  gauge}},  \href{http://dx.doi.org/10.1007/JHEP10(2017)166}{{\em JHEP}
  {\bfseries 10} (2017) 166}, \href{http://arxiv.org/abs/1709.07718}{{\ttfamily
  arXiv:1709.07718 [hep-ph]}}.

\bibitem{Baikov:2014qja}
P.~A. Baikov, K.~G. Chetyrkin, and J.~H. K\"uhn, {\it {Quark Mass and Field
  Anomalous Dimensions to ${\cal O}(\alpha_s^5)$}},
  \href{http://dx.doi.org/10.1007/JHEP10(2014)076}{{\em JHEP} {\bfseries 10}
  (2014) 076}, \href{http://arxiv.org/abs/1402.6611}{{\ttfamily arXiv:1402.6611
  [hep-ph]}}.

\bibitem{Luthe:2016xec}
T.~Luthe, A.~Maier, P.~Marquard, and Y.~Schr\"oder, {\it {Five-loop quark mass
  and field anomalous dimensions for a general gauge group}},
  \href{http://dx.doi.org/10.1007/JHEP01(2017)081}{{\em JHEP} {\bfseries 01}
  (2017) 081}, \href{http://arxiv.org/abs/1612.05512}{{\ttfamily
  arXiv:1612.05512 [hep-ph]}}.

\bibitem{Baikov:2017ujl}
P.~A. Baikov, K.~G. Chetyrkin, and J.~H. K\"uhn, {\it {Five-loop fermion
  anomalous dimension for a general gauge group from four-loop massless
  propagators}},  \href{http://dx.doi.org/10.1007/JHEP04(2017)119}{{\em JHEP}
  {\bfseries 04} (2017) 119}, \href{http://arxiv.org/abs/1702.01458}{{\ttfamily
  arXiv:1702.01458 [hep-ph]}}.

\bibitem{Jegerlehner:2000dz}
F.~Jegerlehner, {\it {Facts of life with $\gamma_5$}},
  \href{http://dx.doi.org/10.1007/s100520100573}{{\em Eur. Phys. J.} {\bfseries
  C18} (2001) 673--679},
\href{http://arxiv.org/abs/hep-th/0005255}{{\ttfamily arXiv:hep-th/0005255
  [hep-th]}}.

\bibitem{Adler:1969gk}
S.~L. Adler, {\it {Axial vector vertex in spinor electrodynamics}},
  \href{http://dx.doi.org/10.1103/PhysRev.177.2426}{{\em Phys. Rev.} {\bfseries
  177} (1969) 2426--2438}.

\bibitem{Bell:1969ts}
J.~S. Bell and R.~Jackiw, {\it {A PCAC puzzle: $\pi^0 \to \gamma \gamma$ in the
  $\sigma$ model}},
\href{http://dx.doi.org/10.1007/BF02823296}{{\em Nuovo Cim.} {\bfseries A60}
  (1969) 47--61}.

\bibitem{Bouchiat:1972iq}
C.~Bouchiat, J.~Iliopoulos, and P.~Meyer, {\it {An Anomaly Free Version of
  Weinberg's Model}},
  \href{http://dx.doi.org/10.1016/0370-2693(72)90532-1}{{\em Phys. Lett. B}
  {\bfseries 38} (1972) 519--523}.

\bibitem{Gross:1972pv}
D.~J. Gross and R.~Jackiw, {\it {Effect of anomalies on quasirenormalizable
  theories}},  \href{http://dx.doi.org/10.1103/PhysRevD.6.477}{{\em Phys. Rev.
  D} {\bfseries 6} (1972) 477--493}.

\bibitem{Geng:1989tcu}
C.~Q. Geng and R.~E. Marshak, {\it {Uniqueness of Quark and Lepton
  Representations in the Standard Model From the Anomalies Viewpoint}},
  \href{http://dx.doi.org/10.1103/PhysRevD.39.693}{{\em Phys. Rev. D}
  {\bfseries 39} (1989) 693}.

\bibitem{Bollini:1972ui}
C.~G. Bollini and J.~J. Giambiagi, {\it {Dimensional Renormalization: The
  Number of Dimensions as a Regularizing Parameter}},
\href{http://dx.doi.org/10.1007/BF02895558}{{\em Nuovo Cim.} {\bfseries B12}
  (1972) 20--26}.

\bibitem{Akyeampong:1973xi}
D.~A. Akyeampong and R.~Delbourgo, {\it {Dimensional regularization, abnormal
  amplitudes and anomalies}},
\href{http://dx.doi.org/10.1007/BF02786835}{{\em Nuovo Cim.} {\bfseries A17}
  (1973) 578--586}.

\bibitem{Breitenlohner:1977hr}
P.~Breitenlohner and D.~Maison, {\it {Dimensional Renormalization and the
  Action Principle}},
\href{http://dx.doi.org/10.1007/BF01609069}{{\em Commun. Math. Phys.}
  {\bfseries 52} (1977) 11--38}.

\bibitem{Breitenlohner:1975hg}
P.~Breitenlohner and D.~Maison, {\it {Dimensionally Renormalized Green's
  Functions for Theories with Massless Particles. 1.}},
  \href{http://dx.doi.org/10.1007/BF01609070}{{\em Commun. Math. Phys.}
  {\bfseries 52} (1977) 39}.

\bibitem{Breitenlohner:1976te}
P.~Breitenlohner and D.~Maison, {\it {Dimensionally Renormalized Green's
  Functions for Theories with Massless Particles. 2.}},
  \href{http://dx.doi.org/10.1007/BF01609071}{{\em Commun. Math. Phys.}
  {\bfseries 52} (1977) 55}.

\bibitem{Bardeen:1972vi}
W.~A. Bardeen, R.~Gastmans, and B.~E. Lautrup, {\it {Static quantities in
  Weinberg's model of weak and electromagnetic interactions}},
\href{http://dx.doi.org/10.1016/0550-3213(72)90218-0}{{\em Nucl. Phys.}
  {\bfseries B46} (1972) 319--331}.

\bibitem{Chanowitz:1979zu}
M.~S. Chanowitz, M.~Furman, and I.~Hinchliffe, {\it {The Axial Current in
  Dimensional Regularization}},
\href{http://dx.doi.org/10.1016/0550-3213(79)90333-X}{{\em Nucl. Phys.}
  {\bfseries B159} (1979) 225--243}.

\bibitem{Gottlieb:1979ix}
S.~A. Gottlieb and J.~T. Donohue, {\it {The Axial Vector Current and
  Dimensional Regularization}},
\href{http://dx.doi.org/10.1103/PhysRevD.20.3378}{{\em Phys. Rev.} {\bfseries
  D20} (1979) 3378}.

\bibitem{Siegel:1979wq}
W.~Siegel, {\it {Supersymmetric Dimensional Regularization via Dimensional
  Reduction}},
\href{http://dx.doi.org/10.1016/0370-2693(79)90282-X}{{\em Phys. Lett.}
  {\bfseries 84B} (1979) 193--196}.

\bibitem{Fujii:1980yt}
Y.~Fujii, N.~Ohta, and H.~Taniguchi, {\it {On the Definitions of $\gamma_5$ in
  Continuous Dimensions}},
  \href{http://dx.doi.org/10.1016/0550-3213(81)90393-X}{{\em Nucl. Phys. B}
  {\bfseries 177} (1981) 297--324}.

\bibitem{Bonneau:1980yb}
G.~Bonneau, {\it {Consistency in Dimensional Regularization With $\gamma_5$}},
  \href{http://dx.doi.org/10.1016/0370-2693(80)90232-4}{{\em Phys. Lett. B}
  {\bfseries 96} (1980) 147--150}.

\bibitem{Ovrut:1981ne}
B.~A. Ovrut, {\it {Axial Vector Ward Identities and Dimensional
  Regularization}},
\href{http://dx.doi.org/10.1016/0550-3213(83)90511-4}{{\em Nucl. Phys.}
  {\bfseries B213} (1983) 241--265}.

\bibitem{Espriu:1982bw}
D.~Espriu and R.~Tarrach, {\it {Renormalization of the Axial Anomaly
  Operators}},
\href{http://dx.doi.org/10.1007/BF01573750}{{\em Z. Phys.} {\bfseries C16}
  (1982) 77}.

\bibitem{Thompson:1985uv}
G.~Thompson and H.~L. Yu, {\it {GAMMA(5) IN DIMENSIONAL REGULARIZATION}},
  \href{http://dx.doi.org/10.1016/0370-2693(85)91397-8}{{\em Phys. Lett. B}
  {\bfseries 151} (1985) 119--122}.

\bibitem{Abdelhafiz:1986jh}
M.~I. Abdelhafiz and M.~Zralek, {\it {The $\gamma$(5) and Dimensional
  Regularization}},  {\em Acta Phys. Polon. B} {\bfseries 18} (1987) 21.

\bibitem{Buras:1989xd}
A.~J. Buras and P.~H. Weisz, {\it {QCD Nonleading Corrections to Weak Decays in
  Dimensional Regularization and 't Hooft-Veltman Schemes}},
\href{http://dx.doi.org/10.1016/0550-3213(90)90223-Z}{{\em Nucl. Phys.}
  {\bfseries B333} (1990) 66--99}.

\bibitem{Kreimer:1989ke}
D.~Kreimer, {\it {The $\gamma_5$ Problem and Anomalies: A Clifford Algebra
  Approach}},
\href{http://dx.doi.org/10.1016/0370-2693(90)90461-E}{{\em Phys. Lett.}
  {\bfseries B237} (1990) 59--62}.

\bibitem{Korner:1991sx}
J.~G. K{\"o}rner, D.~Kreimer, and K.~Schilcher, {\it {A Practicable $\gamma_5$
  scheme in dimensional regularization}},
\href{http://dx.doi.org/10.1007/BF01559471}{{\em Z. Phys.} {\bfseries C54}
  (1992) 503--512}.

\bibitem{Kreimer:1993bh}
D.~Kreimer, {\it {The Role of $\gamma_5$ in Dimensional Regularization}},
  \href{http://arxiv.org/abs/hep-ph/9401354}{{\ttfamily arXiv:hep-ph/9401354}}.

\bibitem{Larin:1991tj}
S.~A. Larin and J.~A.~M. Vermaseren, {\it {The $\alpha_s^3$ corrections to the
  Bjorken sum rule for polarized electroproduction and to the Gross-Llewellyn
  Smith sum rule}},
\href{http://dx.doi.org/10.1016/0370-2693(91)90839-I}{{\em Phys. Lett.}
  {\bfseries B259} (1991) 345--352}.

\bibitem{Larin:1993tq}
S.~A. Larin, {\it {The Renormalization of the axial anomaly in dimensional
  regularization}},  \href{http://dx.doi.org/10.1016/0370-2693(93)90053-K}{{\em
  Phys. Lett.} {\bfseries B303} (1993) 113--118},
\href{http://arxiv.org/abs/hep-ph/9302240}{{\ttfamily arXiv:hep-ph/9302240
  [hep-ph]}}.

\bibitem{Trueman:1995ca}
T.~L. Trueman, {\it {Spurious anomalies in dimensional renormalization}},
  \href{http://dx.doi.org/10.1007/BF02907437}{{\em Z. Phys. C} {\bfseries 69}
  (1996) 525--536}, \href{http://arxiv.org/abs/hep-ph/9504315}{{\ttfamily
  arXiv:hep-ph/9504315}}.

\bibitem{Chetyrkin:1997gb}
K.~G. Chetyrkin, M.~Misiak, and M.~Munz, {\it {$|\Delta F| = 1$ nonleptonic
  effective Hamiltonian in a simpler scheme}},
  \href{http://dx.doi.org/10.1016/S0550-3213(98)00131-X}{{\em Nucl. Phys. B}
  {\bfseries 520} (1998) 279--297},
  \href{http://arxiv.org/abs/hep-ph/9711280}{{\ttfamily arXiv:hep-ph/9711280}}.

\bibitem{Pernici:1999ga}
M.~Pernici, {\it {Seminaive dimensional renormalization}},
  \href{http://dx.doi.org/10.1016/S0550-3213(00)00268-6}{{\em Nucl. Phys. B}
  {\bfseries 582} (2000) 733--755},
  \href{http://arxiv.org/abs/hep-th/9912278}{{\ttfamily arXiv:hep-th/9912278}}.

\bibitem{Wu:2002xa}
Y.-L. Wu, {\it {Symmetry principle preserving and infinity free regularization
  and renormalization of quantum field theories and the mass gap}},
  \href{http://dx.doi.org/10.1142/S0217751X03015222}{{\em Int. J. Mod. Phys. A}
  {\bfseries 18} (2003) 5363--5420},
  \href{http://arxiv.org/abs/hep-th/0209021}{{\ttfamily arXiv:hep-th/0209021}}.

\bibitem{Ma:2005md}
Y.-L. Ma and Y.-L. Wu, {\it {Anomaly and anomaly-free treatment of QFTs based
  on symmetry-preserving loop regularization}},
  \href{http://dx.doi.org/10.1142/S0217751X0603309X}{{\em Int. J. Mod. Phys. A}
  {\bfseries 21} (2006) 6383--6456},
  \href{http://arxiv.org/abs/hep-ph/0509083}{{\ttfamily arXiv:hep-ph/0509083}}.

\bibitem{Tsai:2009it}
E.-C. Tsai, {\it {Gauge Invariant Treatment of $\gamma_{5}$ in the Scheme of 't
  Hooft and Veltman}},
  \href{http://dx.doi.org/10.1103/PhysRevD.83.025020}{{\em Phys. Rev. D}
  {\bfseries 83} (2011) 025020},
  \href{http://arxiv.org/abs/0905.1550}{{\ttfamily arXiv:0905.1550 [hep-th]}}.

\bibitem{Fazio:2014xea}
R.~A. Fazio, P.~Mastrolia, E.~Mirabella, and W.~J. Torres~Bobadilla, {\it {On
  the Four-Dimensional Formulation of Dimensionally Regulated Amplitudes}},
  \href{http://dx.doi.org/10.1140/epjc/s10052-014-3197-4}{{\em Eur. Phys. J. C}
  {\bfseries 74} no.~12, (2014) 3197},
  \href{http://arxiv.org/abs/1404.4783}{{\ttfamily arXiv:1404.4783 [hep-ph]}}.

\bibitem{Moch:2015usa}
S.~Moch, J.~A.~M. Vermaseren, and A.~Vogt, {\it {On $\gamma_5$ in higher-order
  QCD calculations and the NNLO evolution of the polarized valence
  distribution}},  \href{http://dx.doi.org/10.1016/j.physletb.2015.07.027}{{\em
  Phys. Lett.} {\bfseries B748} (2015) 432--438},
\href{http://arxiv.org/abs/1506.04517}{{\ttfamily arXiv:1506.04517 [hep-ph]}}.

\bibitem{Porto:2017asd}
J.~S. Porto, A.~R. Vieira, A.~L. Cherchiglia, M.~Sampaio, and B.~Hiller, {\it
  {On the Bose symmetry and the left- and right-chiral anomalies}},
  \href{http://dx.doi.org/10.1140/epjc/s10052-018-5648-9}{{\em Eur. Phys. J. C}
  {\bfseries 78} no.~2, (2018) 160},
  \href{http://arxiv.org/abs/1706.01001}{{\ttfamily arXiv:1706.01001
  [hep-th]}}.

\bibitem{Bruque:2018bmy}
A.~M. Bruque, A.~L. Cherchiglia, and M.~P\'erez-Victoria, {\it {Dimensional
  regularization vs methods in fixed dimension with and without $\gamma_5$}},
  \href{http://dx.doi.org/10.1007/JHEP08(2018)109}{{\em JHEP} {\bfseries 08}
  (2018) 109}, \href{http://arxiv.org/abs/1803.09764}{{\ttfamily
  arXiv:1803.09764 [hep-ph]}}.

\bibitem{Gnendiger:2017rfh}
C.~Gnendiger and A.~Signer, {\it {$\gamma_{5}$ in the four-dimensional helicity
  scheme}},  \href{http://dx.doi.org/10.1103/PhysRevD.97.096006}{{\em Phys.
  Rev. D} {\bfseries 97} no.~9, (2018) 096006},
  \href{http://arxiv.org/abs/1710.09231}{{\ttfamily arXiv:1710.09231
  [hep-ph]}}.

\bibitem{Zerf:2019ynn}
N.~Zerf, {\it {Fermion Traces Without Evanescence}},
  \href{http://dx.doi.org/10.1103/PhysRevD.101.036002}{{\em Phys. Rev.}
  {\bfseries D101} no.~3, (2020) 036002},
\href{http://arxiv.org/abs/1911.06345}{{\ttfamily arXiv:1911.06345 [hep-ph]}}.

\bibitem{Lang:2021hnw}
J.-N. Lang, S.~Pozzorini, H.~Zhang, and M.~F. Zoller, {\it {Two-loop rational
  terms for spontaneously broken theories}},
  \href{http://dx.doi.org/10.1007/JHEP01(2022)105}{{\em JHEP} {\bfseries 01}
  (2022) 105}, \href{http://arxiv.org/abs/2107.10288}{{\ttfamily
  arXiv:2107.10288 [hep-ph]}}.

\bibitem{Cherchiglia:2021uce}
A.~Cherchiglia, {\it {Step towards a consistent treatment of chiral theories at
  higher loop order: The abelian case}},
  \href{http://dx.doi.org/10.1016/j.nuclphysb.2023.116104}{{\em Nucl. Phys. B}
  {\bfseries 987} (2023) 116104},
  \href{http://arxiv.org/abs/2106.14039}{{\ttfamily arXiv:2106.14039
  [hep-ph]}}.

\bibitem{Rosado:2023ist}
R.~J.~C. Rosado, A.~Cherchiglia, M.~Sampaio, and B.~Hiller, {\it {Infrared
  Subtleties and Chiral Vertices at NLO: An Implicit Regularization Analysis}},
   \href{http://arxiv.org/abs/2305.07129}{{\ttfamily arXiv:2305.07129
  [hep-ph]}}.

\bibitem{OlgosoRuiz:2024dzq}
P.~Olgoso~Ruiz and L.~Vecchi, {\it {Spurious gauge-invariance and $\gamma_5$ in
  Dimensional Regularization}},
  \href{http://arxiv.org/abs/2406.17013}{{\ttfamily arXiv:2406.17013
  [hep-ph]}}.

\bibitem{Draggiotis:2009yb}
P.~Draggiotis, M.~V. Garzelli, C.~G. Papadopoulos, and R.~Pittau, {\it {Feynman
  Rules for the Rational Part of the QCD 1-loop amplitudes}},
  \href{http://dx.doi.org/10.1088/1126-6708/2009/04/072}{{\em JHEP} {\bfseries
  04} (2009) 072}, \href{http://arxiv.org/abs/0903.0356}{{\ttfamily
  arXiv:0903.0356 [hep-ph]}}.

\bibitem{Garzelli:2009is}
M.~V. Garzelli, I.~Malamos, and R.~Pittau, {\it {Feynman rules for the rational
  part of the Electroweak 1-loop amplitudes}},
  \href{http://dx.doi.org/10.1007/JHEP10(2010)097}{{\em JHEP} {\bfseries 01}
  (2010) 040}, \href{http://arxiv.org/abs/0910.3130}{{\ttfamily arXiv:0910.3130
  [hep-ph]}}. [Erratum: JHEP 10, 097 (2010)].

\bibitem{Garzelli:2010qm}
M.~V. Garzelli, I.~Malamos, and R.~Pittau, {\it {Feynman rules for the rational
  part of the Electroweak 1-loop amplitudes in the $R_xi$ gauge and in the
  Unitary gauge}},  \href{http://dx.doi.org/10.1007/JHEP01(2011)029}{{\em JHEP}
  {\bfseries 01} (2011) 029}, \href{http://arxiv.org/abs/1009.4302}{{\ttfamily
  arXiv:1009.4302 [hep-ph]}}.

\bibitem{Shao:2011tg}
H.-S. Shao, Y.-J. Zhang, and K.-T. Chao, {\it {Feynman Rules for the Rational
  Part of the Standard Model One-loop Amplitudes in the 't Hooft-Veltman
  $\gamma_5$ Scheme}},  \href{http://dx.doi.org/10.1007/JHEP09(2011)048}{{\em
  JHEP} {\bfseries 09} (2011) 048},
  \href{http://arxiv.org/abs/1106.5030}{{\ttfamily arXiv:1106.5030 [hep-ph]}}.

\bibitem{Dugan:1990df}
M.~J. Dugan and B.~Grinstein, {\it {On the vanishing of evanescent operators}},
   \href{http://dx.doi.org/10.1016/0370-2693(91)90680-O}{{\em Phys. Lett. B}
  {\bfseries 256} (1991) 239--244}.

\bibitem{Adam:1993uu}
L.~E. Adam and K.~G. Chetyrkin, {\it {Renormalization of four quark operators
  and QCD sum rules}},
  \href{http://dx.doi.org/10.1016/0370-2693(94)90528-2}{{\em Phys. Lett. B}
  {\bfseries 329} (1994) 129--135},
  \href{http://arxiv.org/abs/hep-ph/9404331}{{\ttfamily arXiv:hep-ph/9404331}}.

\bibitem{Herrlich:1994kh}
S.~Herrlich and U.~Nierste, {\it {Evanescent operators, scheme dependences and
  double insertions}},
  \href{http://dx.doi.org/10.1016/0550-3213(95)00474-7}{{\em Nucl. Phys. B}
  {\bfseries 455} (1995) 39--58},
  \href{http://arxiv.org/abs/hep-ph/9412375}{{\ttfamily arXiv:hep-ph/9412375}}.

\bibitem{Wang:2017ijn}
Y.-M. Wang and Y.-L. Shen, {\it {Subleading power corrections to the
  pion-photon transition form factor in QCD}},
  \href{http://dx.doi.org/10.1007/JHEP12(2017)037}{{\em JHEP} {\bfseries 12}
  (2017) 037}, \href{http://arxiv.org/abs/1706.05680}{{\ttfamily
  arXiv:1706.05680 [hep-ph]}}.

\bibitem{Ahmed:2020kme}
T.~Ahmed, W.~Bernreuther, L.~Chen, and M.~Czakon, {\it {Polarized $q \bar{q}
  \rightarrow Z +$Higgs amplitudes at two loops in QCD: the interplay between
  vector and axial vector form factors and a pitfall in applying a
  non-anticommuting $\gamma_5$}},
  \href{http://dx.doi.org/10.1007/JHEP07(2020)159}{{\em JHEP} {\bfseries 07}
  (2020) 159}, \href{http://arxiv.org/abs/2004.13753}{{\ttfamily
  arXiv:2004.13753 [hep-ph]}}.

\bibitem{DiNoi:2023ygk}
S.~Di~Noi, R.~Gr\"ober, G.~Heinrich, J.~Lang, and M.~Vitti, {\it
  {\ensuremath{\gamma}5 schemes and the interplay of SMEFT operators in the
  Higgs-gluon coupling}},
  \href{http://dx.doi.org/10.1103/PhysRevD.109.095024}{{\em Phys. Rev. D}
  {\bfseries 109} no.~9, (2024) 095024},
  \href{http://arxiv.org/abs/2310.18221}{{\ttfamily arXiv:2310.18221
  [hep-ph]}}.

\bibitem{Egner:2024azu}
M.~Egner, M.~Fael, K.~Sch\"onwald, and M.~Steinhauser, {\it {Nonleptonic
  $B$-meson decays to next-to-next-to-leading order}},
  \href{http://arxiv.org/abs/2406.19456}{{\ttfamily arXiv:2406.19456
  [hep-ph]}}.

\bibitem{Bern:1994zx}
Z.~Bern, L.~J. Dixon, D.~C. Dunbar, and D.~A. Kosower, {\it {One loop n point
  gauge theory amplitudes, unitarity and collinear limits}},
  \href{http://dx.doi.org/10.1016/0550-3213(94)90179-1}{{\em Nucl. Phys.}
  {\bfseries B425} (1994) 217--260},
\href{http://arxiv.org/abs/hep-ph/9403226}{{\ttfamily arXiv:hep-ph/9403226
  [hep-ph]}}.

\bibitem{Bern:1994cg}
Z.~Bern, L.~J. Dixon, D.~C. Dunbar, and D.~A. Kosower, {\it {Fusing gauge
  theory tree amplitudes into loop amplitudes}},
  \href{http://dx.doi.org/10.1016/0550-3213(94)00488-Z}{{\em Nucl. Phys.}
  {\bfseries B435} (1995) 59--101},
\href{http://arxiv.org/abs/hep-ph/9409265}{{\ttfamily arXiv:hep-ph/9409265
  [hep-ph]}}.

\bibitem{Bern:1997sc}
Z.~Bern, L.~J. Dixon, and D.~A. Kosower, {\it {One loop amplitudes for e+ e- to
  four partons}},  \href{http://dx.doi.org/10.1016/S0550-3213(97)00703-7}{{\em
  Nucl. Phys.} {\bfseries B513} (1998) 3--86},
\href{http://arxiv.org/abs/hep-ph/9708239}{{\ttfamily arXiv:hep-ph/9708239
  [hep-ph]}}.

\bibitem{Britto:2004nc}
R.~Britto, F.~Cachazo, and B.~Feng, {\it {Generalized unitarity and one-loop
  amplitudes in N=4 super-Yang-Mills}},
  \href{http://dx.doi.org/10.1016/j.nuclphysb.2005.07.014}{{\em Nucl. Phys.}
  {\bfseries B725} (2005) 275--305},
\href{http://arxiv.org/abs/hep-th/0412103}{{\ttfamily arXiv:hep-th/0412103
  [hep-th]}}.

\bibitem{Bern:2007dw}
Z.~Bern, L.~J. Dixon, and D.~A. Kosower, {\it {On-Shell Methods in Perturbative
  QCD}},  \href{http://dx.doi.org/10.1016/j.aop.2007.04.014}{{\em Annals Phys.}
  {\bfseries 322} (2007) 1587--1634},
\href{http://arxiv.org/abs/0704.2798}{{\ttfamily arXiv:0704.2798 [hep-ph]}}.

\bibitem{Ossola:2006us}
G.~Ossola, C.~G. Papadopoulos, and R.~Pittau, {\it {Reducing full one-loop
  amplitudes to scalar integrals at the integrand level}},
  \href{http://dx.doi.org/10.1016/j.nuclphysb.2006.11.012}{{\em Nucl. Phys. B}
  {\bfseries 763} (2007) 147--169},
  \href{http://arxiv.org/abs/hep-ph/0609007}{{\ttfamily arXiv:hep-ph/0609007}}.

\bibitem{Ellis:2007br}
R.~K. Ellis, W.~T. Giele, and Z.~Kunszt, {\it {A Numerical Unitarity Formalism
  for Evaluating One-Loop Amplitudes}},
  \href{http://dx.doi.org/10.1088/1126-6708/2008/03/003}{{\em JHEP} {\bfseries
  03} (2008) 003}, \href{http://arxiv.org/abs/0708.2398}{{\ttfamily
  arXiv:0708.2398 [hep-ph]}}.

\bibitem{Trueman:1979en}
T.~L. Trueman, {\it {Chiral Symmetry in Perturbative {QCD}}},
\href{http://dx.doi.org/10.1016/0370-2693(79)90480-5}{{\em Phys. Lett.}
  {\bfseries 88B} (1979) 331--334}.

\bibitem{Bos:1992nd}
M.~Bos, {\it {Explicit calculation of the renormalized singlet axial anomaly}},
   \href{http://dx.doi.org/10.1016/0550-3213(93)90479-9}{{\em Nucl. Phys. B}
  {\bfseries 404} (1993) 215--244},
  \href{http://arxiv.org/abs/hep-ph/9211319}{{\ttfamily arXiv:hep-ph/9211319}}.

\bibitem{Larin:1997qq}
S.~A. Larin, T.~van Ritbergen, and J.~A.~M. Vermaseren, {\it {The $\alpha_s^3$
  approximation of quantum chromodynamics to the Ellis-Jaffe sum rule}},
  \href{http://dx.doi.org/10.1016/S0370-2693(97)00534-0}{{\em Phys. Lett. B}
  {\bfseries 404} (1997) 153--160},
  \href{http://arxiv.org/abs/hep-ph/9702435}{{\ttfamily arXiv:hep-ph/9702435}}.

\bibitem{Rittinger:2012the}
J.~Rittinger, \href{http://dx.doi.org/10.5445/IR/1000028694}{{\em {Totale
  Zerfallsrate des Z-Bosons nach Hadronen zur Ordnung $\alpha_s^4$}}}.
\newblock PhD thesis, KIT, Karlsruhe, 2012.

\bibitem{Ahmed:2021spj}
T.~Ahmed, L.~Chen, and M.~Czakon, {\it {Renormalization of the flavor-singlet
  axial-vector current and its anomaly in dimensional regularization}},
  \href{http://dx.doi.org/10.1007/JHEP05(2021)087}{{\em JHEP} {\bfseries 05}
  (2021) 087}, \href{http://arxiv.org/abs/2101.09479}{{\ttfamily
  arXiv:2101.09479 [hep-ph]}}.

\bibitem{Chen:2021gxv}
L.~Chen and M.~Czakon, {\it {Renormalization of the axial current operator in
  dimensional regularization at four-loop in QCD}},
  \href{http://dx.doi.org/10.1007/JHEP01(2022)187}{{\em JHEP} {\bfseries 01}
  (2022) 187}, \href{http://arxiv.org/abs/2112.03795}{{\ttfamily
  arXiv:2112.03795 [hep-ph]}}.

\bibitem{Chen:2022lun}
L.~Chen and M.~Czakon, {\it {The $\overline{\mathrm{MS}}$ renormalization
  constant of the singlet axial current operator at $\mathcal{O}(\alpha_s^5)$
  in QCD}},  \href{http://dx.doi.org/10.1016/j.physletb.2022.137266}{{\em Phys.
  Lett. B} {\bfseries 832} (2022) 137266},
  \href{http://arxiv.org/abs/2201.01797}{{\ttfamily arXiv:2201.01797
  [hep-ph]}}.

\bibitem{Martin:1999cc}
C.~P. Martin and D.~Sanchez-Ruiz, {\it {Action principles, restoration of BRS
  symmetry and the renormalization group equation for chiral nonAbelian gauge
  theories in dimensional renormalization with a nonanticommuting gamma(5)}},
  \href{http://dx.doi.org/10.1016/S0550-3213(99)00453-8}{{\em Nucl. Phys. B}
  {\bfseries 572} (2000) 387--477},
  \href{http://arxiv.org/abs/hep-th/9905076}{{\ttfamily arXiv:hep-th/9905076}}.

\bibitem{Grassi:1999tp}
P.~A. Grassi, T.~Hurth, and M.~Steinhauser, {\it {Practical algebraic
  renormalization}},  \href{http://dx.doi.org/10.1006/aphy.2001.6117}{{\em
  Annals Phys.} {\bfseries 288} (2001) 197--248},
  \href{http://arxiv.org/abs/hep-ph/9907426}{{\ttfamily arXiv:hep-ph/9907426}}.

\bibitem{Sanchez-Ruiz:2002pcf}
D.~Sanchez-Ruiz, {\it {BRS symmetry restoration of chiral Abelian Higgs-Kibble
  theory in dimensional renormalization with a nonanticommuting gamma(5)}},
  \href{http://dx.doi.org/10.1103/PhysRevD.68.025009}{{\em Phys. Rev. D}
  {\bfseries 68} (2003) 025009},
  \href{http://arxiv.org/abs/hep-th/0209023}{{\ttfamily arXiv:hep-th/0209023}}.

\bibitem{Cornella:2022hkc}
C.~Cornella, F.~Feruglio, and L.~Vecchi, {\it {Gauge invariance and finite
  counterterms in chiral gauge theories}},
  \href{http://dx.doi.org/10.1007/JHEP02(2023)244}{{\em JHEP} {\bfseries 02}
  (2023) 244}, \href{http://arxiv.org/abs/2205.10381}{{\ttfamily
  arXiv:2205.10381 [hep-ph]}}.

\bibitem{Belusca-Maito:2020ala}
H.~B\'elusca-Ma\"\i{}to, A.~Ilakovac, M.~Ma\dj{}or-Bo\v{z}inovi\'c, and
  D.~St\"ockinger, {\it {Dimensional regularization and
  Breitenlohner-Maison/\textquoteright{}t Hooft-Veltman scheme for $\gamma_5$
  applied to chiral YM theories: full one-loop counterterm and RGE structure}},
   \href{http://dx.doi.org/10.1007/JHEP08(2020)024}{{\em JHEP} {\bfseries 08}
  no.~08, (2020) 024}, \href{http://arxiv.org/abs/2004.14398}{{\ttfamily
  arXiv:2004.14398 [hep-ph]}}.

\bibitem{Belusca-Maito:2021lnk}
H.~B\'elusca-Ma\"\i{}to, A.~Ilakovac, P.~K\"uhler,
  M.~Ma\dj{}or-Bo\v{z}inovi\'c, and D.~St\"ockinger, {\it {Two-loop application
  of the Breitenlohner-Maison/\textquoteright{}t Hooft-Veltman scheme with
  non-anticommuting \ensuremath{\gamma}$_{5}$: full renormalization and
  symmetry-restoring counterterms in an abelian chiral gauge theory}},
  \href{http://dx.doi.org/10.1007/JHEP11(2021)159}{{\em JHEP} {\bfseries 11}
  (2021) 159}, \href{http://arxiv.org/abs/2109.11042}{{\ttfamily
  arXiv:2109.11042 [hep-ph]}}.

\bibitem{Stockinger:2023ndm}
D.~St\"ockinger and M.~Wei\ss{}wange, {\it {Full three-loop renormalisation of
  an abelian chiral gauge theory with non-anticommuting
  \ensuremath{\gamma}$_{5}$ in the BMHV scheme}},
  \href{http://dx.doi.org/10.1007/JHEP02(2024)139}{{\em JHEP} {\bfseries 02}
  (2024) 139}, \href{http://arxiv.org/abs/2312.11291}{{\ttfamily
  arXiv:2312.11291 [hep-ph]}}.

\bibitem{Chen:2023lus}
L.~Chen, {\it {An observation on Feynman diagrams with axial anomalous
  subgraphs in dimensional regularization with an anticommuting
  \ensuremath{\gamma}$_{5}$}},
  \href{http://dx.doi.org/10.1007/JHEP11(2023)030}{{\em JHEP} {\bfseries 2023}
  no.~11, (2023) 30}, \href{http://arxiv.org/abs/2304.13814}{{\ttfamily
  arXiv:2304.13814 [hep-ph]}}.

\bibitem{Adler:1969er}
S.~L. Adler and W.~A. Bardeen, {\it {Absence of higher order corrections in the
  anomalous axial vector divergence equation}},
\href{http://dx.doi.org/10.1103/PhysRev.182.1517}{{\em Phys. Rev.} {\bfseries
  182} (1969) 1517--1536}.

\bibitem{Collins:1984xc}
J.~C. Collins, \href{http://dx.doi.org/10.1017/CBO9780511622656}{{\em
  {Renormalization}}}, vol.~26 of {\em Cambridge Monographs on Mathematical
  Physics}.
\newblock Cambridge University Press, Cambridge,
1986.
\newblock

\bibitem{Nogueira:1991ex}
P.~Nogueira, {\it {Automatic Feynman graph generation}},
\href{http://dx.doi.org/10.1006/jcph.1993.1074}{{\em J. Comput. Phys.}
  {\bfseries 105} (1993) 279--289}.

\bibitem{Poole:2019txl}
C.~Poole and A.~E. Thomsen, {\it {Weyl Consistency Conditions and
  \ensuremath{\gamma}$_5$}},
  \href{http://dx.doi.org/10.1103/PhysRevLett.123.041602}{{\em Phys. Rev.
  Lett.} {\bfseries 123} no.~4, (2019) 041602},
  \href{http://arxiv.org/abs/1901.02749}{{\ttfamily arXiv:1901.02749
  [hep-th]}}.

\bibitem{Lee:1977eg}
B.~W. Lee, C.~Quigg, and H.~B. Thacker, {\it {Weak Interactions at Very
  High-Energies: The Role of the Higgs Boson Mass}},
  \href{http://dx.doi.org/10.1103/PhysRevD.16.1519}{{\em Phys. Rev. D}
  {\bfseries 16} (1977) 1519}.

\bibitem{Chanowitz:1985hj}
M.~S. Chanowitz and M.~K. Gaillard, {\it {The TeV Physics of Strongly
  Interacting W's and Z's}},
  \href{http://dx.doi.org/10.1016/0550-3213(85)90580-2}{{\em Nucl. Phys. B}
  {\bfseries 261} (1985) 379--431}.

\bibitem{Cornwall:1974km}
J.~M. Cornwall, D.~N. Levin, and G.~Tiktopoulos, {\it {Derivation of Gauge
  Invariance from High-Energy Unitarity Bounds on the s Matrix}},
  \href{http://dx.doi.org/10.1103/PhysRevD.10.1145}{{\em Phys. Rev. D}
  {\bfseries 10} (1974) 1145}. [Erratum: Phys.Rev.D 11, 972 (1975)].

\bibitem{Vayonakis:1976vz}
C.~E. Vayonakis, {\it {Born Helicity Amplitudes and Cross-Sections in
  Nonabelian Gauge Theories}},
  \href{http://dx.doi.org/10.1007/BF02746538}{{\em Lett. Nuovo Cim.} {\bfseries
  17} (1976) 383}.

\bibitem{Chetyrkin:2012rz}
K.~G. Chetyrkin and M.~F. Zoller, {\it {Three-loop
  \textbackslash{}beta-functions for top-Yukawa and the Higgs self-interaction
  in the Standard Model}},
  \href{http://dx.doi.org/10.1007/JHEP06(2012)033}{{\em JHEP} {\bfseries 06}
  (2012) 033}, \href{http://arxiv.org/abs/1205.2892}{{\ttfamily arXiv:1205.2892
  [hep-ph]}}.

\bibitem{Siegel:1980qs}
W.~Siegel, {\it {Inconsistency of Supersymmetric Dimensional Regularization}},
  \href{http://dx.doi.org/10.1016/0370-2693(80)90819-9}{{\em Phys. Lett. B}
  {\bfseries 94} (1980) 37--40}.

\bibitem{Gnendiger:2017pys}
C.~Gnendiger {\em et~al.}, {\it {To ${d}$, or not to ${d}$: recent developments
  and comparisons of regularization schemes}},
  \href{http://dx.doi.org/10.1140/epjc/s10052-017-5023-2}{{\em Eur. Phys. J. C}
  {\bfseries 77} no.~7, (2017) 471},
  \href{http://arxiv.org/abs/1705.01827}{{\ttfamily arXiv:1705.01827
  [hep-ph]}}.

\bibitem{Kinoshita:1962ur}
T.~Kinoshita, {\it {Mass singularities of Feynman amplitudes}},
  \href{http://dx.doi.org/10.1063/1.1724268}{{\em J. Math. Phys.} {\bfseries 3}
  (1962) 650--677}.

\bibitem{Lee:1964is}
T.~D. Lee and M.~Nauenberg, {\it {Degenerate Systems and Mass Singularities}},
  \href{http://dx.doi.org/10.1103/PhysRev.133.B1549}{{\em Phys. Rev.}
  {\bfseries 133} (1964) B1549--B1562}.

\bibitem{Cutkosky:1960sp}
R.~E. Cutkosky, {\it {Singularities and discontinuities of Feynman
  amplitudes}},  \href{http://dx.doi.org/10.1063/1.1703676}{{\em J. Math.
  Phys.} {\bfseries 1} (1960) 429--433}.

\bibitem{Chen:2019wyb}
L.~Chen, {\it {A prescription for projectors to compute helicity amplitudes in
  D dimensions}},
  \href{http://dx.doi.org/10.1140/epjc/s10052-021-09210-9}{{\em Eur. Phys. J.
  C} {\bfseries 81} no.~5, (2021) 417},
  \href{http://arxiv.org/abs/1904.00705}{{\ttfamily arXiv:1904.00705
  [hep-ph]}}.

\bibitem{Catani:1999nf}
S.~Catani and M.~H. Seymour, {\it {Corrections of $\mathcal{O}(\alpha_s^2)$ to
  the forward backward asymmetry}},
  \href{http://dx.doi.org/10.1088/1126-6708/1999/07/023}{{\em JHEP} {\bfseries
  07} (1999) 023}, \href{http://arxiv.org/abs/hep-ph/9905424}{{\ttfamily
  arXiv:hep-ph/9905424}}.

\bibitem{Vermaseren:2000nd}
J.~A.~M. Vermaseren, {\it {New features of FORM}},
\href{http://arxiv.org/abs/math-ph/0010025}{{\ttfamily arXiv:math-ph/0010025
  [math-ph]}}.

\bibitem{Mertig:1990an}
R.~Mertig, M.~Bohm, and A.~Denner, {\it {FEYN CALC: Computer algebraic
  calculation of Feynman amplitudes}},
  \href{http://dx.doi.org/10.1016/0010-4655(91)90130-D}{{\em Comput. Phys.
  Commun.} {\bfseries 64} (1991) 345--359}.

\bibitem{Shtabovenko:2016sxi}
V.~Shtabovenko, R.~Mertig, and F.~Orellana, {\it {New Developments in FeynCalc
  9.0}},  \href{http://dx.doi.org/10.1016/j.cpc.2016.06.008}{{\em Comput. Phys.
  Commun.} {\bfseries 207} (2016) 432--444},
  \href{http://arxiv.org/abs/1601.01167}{{\ttfamily arXiv:1601.01167
  [hep-ph]}}.

\bibitem{Shtabovenko:2023idz}
V.~Shtabovenko, R.~Mertig, and F.~Orellana, {\it {FeynCalc 10: Do multiloop
  integrals dream of computer codes?}},
  \href{http://dx.doi.org/10.1016/j.cpc.2024.109357}{{\em Comput. Phys.
  Commun.} {\bfseries 306} (2025) 109357},
  \href{http://arxiv.org/abs/2312.14089}{{\ttfamily arXiv:2312.14089
  [hep-ph]}}.

\bibitem{Patel:2015tea}
H.~H. Patel, {\it {Package-X: A Mathematica package for the analytic
  calculation of one-loop integrals}},
  \href{http://dx.doi.org/10.1016/j.cpc.2015.08.017}{{\em Comput. Phys.
  Commun.} {\bfseries 197} (2015) 276--290},
\href{http://arxiv.org/abs/1503.01469}{{\ttfamily arXiv:1503.01469 [hep-ph]}}.

\bibitem{calcloop}
Y.-Q. Ma, {\it {CalcLoop}},
  \href{http://dx.doi.org/https://gitlab.com/multiloop-pku/calcloop}{{\em
  https://gitlab.com/multiloop-pku/calcloop} }.

\end{thebibliography}\endgroup
\bibliographystyle{utphysM}
\end{document}